\documentclass[twocolumn,trackchanges]{aastex63}
\usepackage{amsmath}
\usepackage{graphicx}
\usepackage{soul}

\newcommand{\ramses}{\texttt{RAMSES}}
\newcommand{\music}{\texttt{MUSIC}}

\newcommand{\camb}{\texttt{CAMB}}
\newcommand{\msun}{\rm M_{\odot}}
\newcommand{\ysamtm}{\texttt{ySAMtm}}
\newcommand{\pgalf}{\texttt{PGalF}}

\newcommand{\hr}{\texttt{HR5}}

\newcommand{\psb}{\texttt{PSB}}

\usepackage{soul} % To allow for the use of \hl.

\shorttitle{Formation and Morphology of the First Galaxies}
\shortauthors{Park et al.}

\begin{document}
\title{Formation and Morphology of the First Galaxies in the Cosmic Morning }

\author[0000-0001-9521-6397]{Changbom Park}
\affiliation{Korea Institute for Advanced Study, 85 Hoegi-ro, Dongdaemun-gu, Seoul 02455, Korea}

\author[0000-0002-6810-1778]{Jaehyun Lee}
\affiliation{Korea Institute for Advanced Study, 85 Hoegi-ro, Dongdaemun-gu, Seoul 02455, Republic of Korea}

\author[0000-0002-4391-2275]{Juhan Kim}
\affiliation{Korea Institute for Advanced Study, 85 Hoegi-ro, Dongdaemun-gu, Seoul 02455, Republic of Korea}

\author{Donghui Jeong}
\affiliation{The Pennsylvania State University, University Park, PA, 16802, USA}
\affiliation{Korea Institute for Advanced Study, 85 Hoegi-ro, Dongdaemun-gu, Seoul 02455, Republic of Korea}

\author[0000-0003-0695-6735]{Christophe Pichon}
\affiliation{CNRS and Sorbonne Universit\'e, UMR 7095, Institut d'Astrophysique de Paris, 98 bis, Boulevard Arago, F-75014 Paris, France}
\affiliation{ IPhT, DRF-INP, UMR 3680, CEA, L'Orme des Merisiers, B\^at 774, 91191 Gif-sur-Yvette, France}
\affiliation{Korea Institute for Advanced Study, 85 Hoegi-ro, Dongdaemun-gu, Seoul 02455, Republic of Korea}

\author[0000-0003-4446-3130]{Brad K. Gibson}
\affiliation{E.A. Milne Centre for Astrophysics, University of Hull, Hull, HU6 7RX, United Kingdom}

\author{Owain N. Snaith}
\affiliation{GEPI, Observatoire de Paris, PSL Research University, CNRS, 5 Place Jules Janssen, 92190, Meudon, France}

\author[0000-0001-5135-1693]{Jihye Shin}
\affiliation{Korea Astronomy and Space Science Institute, 776 Daedeokdae-ro, Yuseong-gu, Daejeon 34055, Korea}

\author{Yonghwi Kim}
\affiliation{Korea Institute for Advanced Study, 85 Hoegi-ro, Dongdaemun-gu, Seoul 02455, Republic of Korea}

\author[0000-0003-0225-6387]{Yohan Dubois}
\affiliation{CNRS and Sorbonne Universit\'e, UMR 7095, Institut d'Astrophysique de Paris, 98 bis, Boulevard Arago, F-75014 Paris, France}

\author{C. Gareth Few}
\affiliation{E.A. Milne Centre for Astrophysics, University of Hull, Hull, HU6 7RX, United Kingdom}

\email{Corresponding author: Jaehyun Lee (syncphy@gmail.com)}

\begin{abstract}
We investigate the formation and morphological evolution of the first galaxies in the cosmic morning ($10 \gtrsim z \gtrsim 4$) using the Horizon Run 5 (\hr) simulation. For galaxies above the stellar mass $M_{\star, {\rm min}} = 2\times 10^9\,\msun$, we classify them into disk, spheroid, and irregular types according to their asymmetry and stellar mass morphology. We find that about 2/3 of the galaxies have a S\'{e}rsic index $< 1.5$, reflecting the dominance of disk-type morphology in the cosmic morning. The rest are evenly distributed as incidental and transient irregulars or spheroids. These fractions are roughly independent of redshift and stellar mass up to $\sim10^{10}\,\msun$. Almost all the first galaxies with $M_{\star}> M_{\star, {\rm min}}$ at $z>4$ form at initial peaks of the matter density field. Large-scale structures in the universe emerge and grow like cosmic rhizomes as the underlying matter density fluctuations grow and form associations of galaxies in rare overdense regions and the realm of the galactic world is stretched into relatively lower-density regions along evolving filaments. The cosmic web of galaxies forms at lower redshifts when most rhizomes globally percolate. The primordial angular momentum produced by the induced tidal torques on protogalactic regions is correlated with the internal kinematics of galaxies and tightly aligned with the angular momentum of the total galaxy mass. The large-scale tidal field imprinted in the initial conditions seems responsible for the dominance of disk morphology, and for the tendency of galaxies to re-acquire a disk post-distortion.

\end{abstract}

%{\TK amount} of \underline{{\JH the stripped ISM plays a critical}} role in characterizing jellyfish galaxies {\JH undergoing strong ram pressure}\footnote{Here you should be careful because strong RAM pressure can always strip the ISM. Please re-write this sentence.}.

%Based on our experiments and observed cases, we suggest that the observed galaxies with massive molecular clouds in their RPS tails probably had extremely gas-rich disks with abundant dusts before falling into clusters.

%Our results suggest that a gas rich galaxy can form characteristic jellyfish features when its ISM is largely stripped by strong ram pressure, mixes with the ICM, helping the cooling of tail clouds.}


%https://www.overleaf.com/project/6193aabba73ed328a956385c
\section {Introduction}

Which morphology do galaxies have when they form: do the first galaxies have a single morphology or an initial morphological distribution? 
How do the initial conditions govern the formation and morphology of the first galaxies? How do  subsequent evolutionary processes affect galaxy morphology? 
Galaxy morphology is one of the key physical properties of galaxies, but these questions regarding the galaxy morphology have not yet been fully answered. At the present epoch, the morphology of most bright galaxies can be categorized according to Hubble's tuning fork~\citep{hubble26}. In this system,
galaxies are classified into irregular  and regular types, which are in turn divided into ellipticals, spirals, and lenticulars.  

Using a volume-limited sample of 1332 GOODS galaxies with rest-frame $M_B < -18$ and $0.4 \leq z \leq 1.0$, %$B$-band absolute magnitude brighter than $-18.0$ 
\citet{hwang09} have found that the fraction of early-type galaxies (ellipticals plus lenticulars) has been lower in the past, reaching as low as about 15\% near $z=1$. They have also found the morphology-density relation~\citep{oemler74,dressler80,park07} observed at low redshifts almost disappears near $z=1$. Namely, near $z=1$, the early-type fraction stays at $\sim$15\% regardless of the density environment.
Similarly, \citet{tamburri14} used 1302 GOODS-MUSIC galaxies brighter than $Ks({\rm AB})=22$ and with stellar mass roughly between $10^{10}$ and $3\times 10^{11}M_{\odot}$ to find that at redshift $z\approx 1$ late types are dominant and the early-type galaxy fraction is about 0.19. A natural question raised from these findings is whether or not galaxies at much higher redshifts ($z\gg 1$) are all late-type galaxies such as disks and irregulars.

There have been many observational efforts to study galaxy morphology at higher redshifts~\citep[see][for reviews]{elmegreen15,marshall19}. However, a consensus is yet to be reached.
%{\JH It seems that Marshall et al  studied high z morphology using semi-analytic models, not observations. }. 
Some observations support the claim that high-redshift galaxies are more clumpy and irregular than low-redshift galaxies~\citep{abraham01,papovich05,elmegreen14}. \citet{cameron11} have reported using the Hubble Ultra Deep Field data and the GOODS South data that irregulars are the most prominent type at $z>2$. On the other hand, many reports do not support the above claims, and find a larger fraction of spheroidals  at high redshifts~\citep{bundy05,franceschini06,lotz06,dahlen07}. 

\citet{ravindranath06} have reported using GOODS images that, at $z\sim 3$, the 40\% of Lyman break galaxies have exponential profiles, 30\% have de Vaucouleurs profiles, and about 30\% have disturbed morphology. For massive galaxies with $M_{\star} > 10^{11}\,\msun$ \citet{buitrago13} have found using a sample of HST galaxies that the fraction of peculiar-type galaxies is about the same as that of early- and late-type galaxies at $z \approx 2$.  The recently launched James Webb space telescope will keep a close eye on this high-redshift universe and reveal the morphology of high-redshift galaxies.

Cosmological simulations have been invaluable tools to understand the formation and evolution of cosmic structures~\citep[e.g.,][]{park90,Ocvirk2008M,park12,dubois14b,genel14,Khandai15,schaye15,Dolag16,tremmel17,mccarthy17,marinacci18,dave19,lee21}. In recent years, several studies using large cosmological simulations accounting for gravity, hydrodynamics, and various sub-grid physics have addressed galaxy morphology at high redshifts and the origin of Hubble's tuning fork classification system.

\cite{Dekel2009} have highlighted the specific importance of cold  streams  as the main mode of gas acquisition at high redshifts, while \cite{Pichon2011} and \cite{Stewart2013}
have shown how steady angular momentum advected along these streams are key to the disk reformation. The consistency of vorticity-rich inflow~\citep{Laigle2015} is due to the coherency of tides on larger scales~\citep{Danovich2015}, which are themselves a result of the structure of the initial conditions~\citep{Bond1996}.
Because cold gas undergoes shocks during shell crossing~\citep{Keres2005}, it closely follows the ridges in the large-scale structure.

From the Horizon simulation, \citet{Dubois2016} highlights the interplay between cosmic gas accretion onto galaxies and galaxy mergers in driving the observed diversity in galaxy morphology at lower redshifts~\citep[see also][]{Martin2018}. They have noted the critical role of active galactic nuclei in setting up the correct galaxy morphology for the massive end of the population. AGN accomplish this by freezing the morphological type of the post-merger remnant, and quenching its quiescent star formation for a considerable duration.

\citet{pillepich19} have used the Illustris TNG50 simulations to focus on the evolution of star-forming galaxies and conclude that the dominant galaxy morphology at $z>2$ is spheroidal for galaxies with $M_{\star} < 10^{10}\,\msun$. They also claim that the fraction of disk-like galaxies increases at lower redshifts, and for larger stellar mass. According to their results, disk-type galaxies become dominant at $z<2$ for actively star-forming galaxies with $M_{\star} > 10^{9.5}\,\msun$. 

\citet{trayford19} have analyzed a sample of simulated galaxies with $M_{\star} > 10^{9}\,\msun$ in the EAGLE simulation. They assign morphology to galaxies using the fraction of the stellar mass that lies in the disk component, and claimed that, at  redshifts $z \gtrsim 1.5$, asymmetric galaxies dominate for all masses. They also reported that the Hubble sequence is established by $z \simeq 1.5$, and the spheroid type dominates after that epoch. The disk type is always subdominant in this work. It should be pointed out that their morphology classification of disk and spheroid types is
based not on the shape of mass distribution
but actually on kinematics parameterized by the circularity $\epsilon$.

\citet{park19} used the New Horizon simulation to find that thin disks in galaxies start to form only at $z\sim1-2$ for those with $M_{\star} (z=0.7) > 10^{10}\,\msun$ or at $z \approx 1$ for $M_{\star} (z=0.7) = 10^9- 10^{10}\,\msun$. The former era 
%for the formation of the thin disk 
matches well with some measurements of the history of the Milky Way~\citep[e.g.][]{2013A&A...560A.109H}.

Hence, neither observations nor simulations give a clear and consistent picture of the initial morphology of high-redshift galaxies. In this study, we use the recent Horizon Run 5 cosmological simulation~\citep{lee21}, \hr, to inspect galaxies at high redshifts $10 \gtrsim z \gtrsim 4$ that we call {\it cosmic morning}%
\footnote{We shall call this period `the cosmic morning,' reserving the name `cosmic dawn' for the period before $z\simeq 10$. Similarly, the cosmic noon is from $z \simeq 4$ to 1.5, and the cosmic afternoon is the period of $z \lesssim 1.5$.}. We will study the physical properties of the first galaxies and large-scale structures in the cosmic morning focusing on their morphology. Relying on the unique volume and resolution of \hr, we shall clarify the origin and early evolution of the morphology of galaxies, respectively, by studying the relationship between  morphology and the initial conditions and by closely examining the cosmic evolution of physical parameters.

\section{Simulation Data}
The cosmological simulation \hr~\citep{lee21} covers a volume of the universe in an 1.15\,cGpc$^3$ scale volume while achieving a resolution down to $\sim$1 pkpc. We adopt the cosmological parameters of $\Omega_{\rm m}=0.3$, $\Omega_{\Lambda}=0.7$, $\Omega_{\rm b}=0.047$, $\sigma_8=0.816$, and $h=0.684$ that are compatible with the Planck data~\citep{planck16}. The linear power spectrum is computed using the \camb\ package~\citep{lewis00} and the initial conditions are generated at $z=200$ using the \music\ package~\citep{hahn11} with second-order Lagrangian perturbation theory~\citep[2LPT;][]{scoccimarro98,lhuillier14}. 

We use a modified version of the adaptive mesh refinement code \ramses~\citep{teyssier02} for \hr. The Poisson equations and Euler equations are solved based on the Particle-Mesh method~\citep{guillet11} and the Harten-Lax-van Leer-contact wave Riemann solver ~\citep{toro94}, respectively. All particles of dark matter, stars, and massive black holes (MBHs) are assigned into grids using a cloud-in-cell scheme to construct mass density grids. We have implemented OpenMP parallelism into \ramses\, on top of the Message Passing Interface originally equipped in \ramses. 

We set a high-resolution cuboid region of $1048.6\times119.0\times127.2\,{\rm cMpc}^3$ at the center of the simulation box of $1048.6^3$\,cMpc$^3$. The cosmological simulation box is set to have a background grid level 8, which corresponds to 256 coarse grids on a side with $\Delta x=4.10\,$cMpc. The high-resolution region is initially filled with $8192\times930\times994$ grids and dark matter particles at a level 13, and the padding grids of levels from 9 to 12 are placed between the coarse and high-resolution regions. A dark matter particle has a mass resolution of $6.89\times10^7\,\msun$ at the highest level, and the particle mass increases by a factor of 8 when the grid level decreases by 1. The grids are adaptively refined down to $\Delta x\sim1\,$pkpc when their mass is larger than eight times the {minimum} dark matter particle mass. We have carried out the \hr\ simulation down to $z=0.625$.

The version of \ramses\ used in \hr\ computes gas cooling down to a temperature of $10^4\,$K using the synthetic cooling functions of \citet{sutherland93}, and metal-enriched gas can additionally cool down to $\sim750\,$K based on the cooling rates proposed by \citet{dalgarno72}. Cosmic reionization is approximated by adopting a uniform UV background of \citet{haardt96}. Star formation is computed using the statistical approach of \citet{rasera06}. The minimum mass of stellar particles is $2.56\times10^6\,\msun$ at birth.

Supernova feedback operates in thermal and kinetic modes~\citep{dubois08} and AGN feedback releases its energy in radio-jet and quasar modes, depending on the Eddington ratio~\citep{dubois12}. Massive black holes (MHBs) are seeded with an initial mass of $10^4\,\msun$ in grids when gas density is higher than the threshold of star formation and no other MBHs exist within 50 kpc~\citep{dubois14b}. MBHs grow via accretion and coalescence, and their angular momentum obtained from the feeding processes are traced~\citep{dubois14a}. We model the chemical evolution by adopting the method proposed by \citet{few12}, and trace the abundance of H, O, and Fe based on a Chabrier initial mass function (IMF)~\citep{chabrier03}. 

We calibrate the subgrid physics of \hr\ to reproduce the star formation rate density, the galaxy stellar mass function, and the mass-metallicity relation using test simulations. The performance of the parameter tuning was assessed by eyes with no formal fitting because of limited computing resources. The details of the calibration are given in Appendix A of \citet{lee21}.

We compute the rest-frame luminosities of stellar particles in the \texttt{Johnson} and \texttt{SDSS} filter systems based on their ages and metallicities during post-processing. We use the photometric predictions generated from \texttt{E-MILES} single stellar population SEDs~\citep{vazdekis12,vazdekis16,ricciardelli12} in this modeling. For consistency with our chemical evolution modeling, we also assume a Chabrier IMF in this calculation. One can find further details of \hr\ in \citet{lee21}.

 %While the  resolution in \hr\, does not reach disk scale height, hence will likely bias morphology estimators, the simulated volume allows us to statistically model the appearance  of the first galaxy population at redshift 10 and below. 

\subsection{Galaxy Finding}
\ramses\ employs four types of astrophysical objects to study their mutual interactions in accordance with gravity and hydrodynamics. Dark matter, stars, and MBHs are represented by point-mass particles while information on gas is assigned to finite-volume cells. The particle-like representation has long been favored in most cosmological $N$-body simulations and various structure-identification methods have successfully been developed using well-known particle-based algorithms, such as the friends of friends (FoF). On the other hand, a gas cell is stationary and has gas quantities being averaged over its volume. In the pipeline of structure finding, we assume the gas cells as particles, like other matter components. Since the particles have varying mass, the linking length of a particle pair is derived by averaging the linking length of each particle.% We therefore, integrate the gas information into the same particle type as adopted for the other matter components.

%To find structures in the \hr, we have implemented the \pgalf\ algorithm (for details, see \citealt{lee21}) based upon the physically-self bound (\texttt{PSB}) subhalo finder~\citep{kjhan06} originally developed for $N$-body simulations~\citep{kjhan15,kjhan09}. The algorithm facilitates the concept of coordinate-free density field and a web of neighboring networks as follows. Instead of counting particles in the cubic grids, we measure mass density at particle positions using the W4 kernel with a fixed number of neighboring particles, where the neighbors of a given particle are defined as its $N$-nearest neighbors. 
%This density field and neighboring networks may form the backbone of the galaxy finding (\pgalf\ hereafter; 
%
To find substructures (or galaxies) in the \hr, we modified the original \psb\ subhalo finder~\citep{kjhan06}, which was developed for $N$-body simulations~\citep{kjhan15,kjhan09}. Like the \psb\ method, \pgalf\ ~\citep{lee21} is based on the coordinate-free density map and a network of neighboring particles. However, \pgalf\ uses the stellar density map if there are stellar particles because stellar distributions are clumpier and more compact, making it easier to determine boundaries of substructures (or galaxies). To identify star particle groups, we set the minimum number of star particles to be ten, which corresponds to $\sim2\times10^7\,\msun$ in the cosmic morning. When \hr\ is calibrated, statistical comparison with observations is made for the `galaxies' with stellar mass above $1\times10^9\,\msun$, and we will use those having $M_{\star} > 2\times10^9\,\msun$ in most parts of our galaxy morphology study. Therefore, galaxies in this study have more than 1,000 star particles.

An unbinding scheme is essential for subhalo identification to minimize contamination by host halo particles~\citep{knebe11}. Thus, various unbinding algorithms have been introduced to halo finders, e.g.,the local escape velocity scheme of AHF~\citep{knollmann09}, the density gradient of SUBFIND~\citep{springel01} or the neighborhood link in 6-dimensional configuration space of Rockstar~\citep{behroozi13} or VELOCIraptor~\citep{elahi19}. Following the scheme of ~\cite{kjhan06} we utilize two criteria to determine the membership of a particle with respect to a substructure.
The first is the tidal radius limit and the other one is the negative total energy criterion. We have used the tidal radius ($r_t$) as
\begin{equation}
    r_t = R \left( {m/M \over \alpha+\beta}\right),
\end{equation}
where $R$ is the distance between the host and satellite galaxies, $M$ is the host galaxy mass contained within radius $R$, 
\begin{equation}
    \alpha(R) \equiv 2 - \left( { d\ln M\over d\ln R}\right)\bigg\rvert_R,
\end{equation}
and
\begin{equation}
    	\beta \equiv {\Omega^2 R^3 \over GM},
\end{equation}
where $\Omega$ is the orbital angular speed of the satellite.
In these equations, $\alpha$ is the contribution of the gravitational force by the host and $\beta$ is the effect of centrifugal force during the orbital motion of the satellite.
%We define the structure-to-structure separation in FoF particle groups by measuring the tidal boundary and total energy of their member particles. 

%The tidal radius of a structure 
%\begin{equation}
% \Sigma(R) = \Sigma_0 {\rm exp} (-k R^{1/n_{\rm S\acute{e}rsic}}), 
% \label{eq:defsersic}
%\end{equation}
A particle is first identified to be bound to a structure when it has negative total energy relative to the structure. The membership is probed if the particle is located inside the tidal radius of the structure.  Therefore, even though a particle has negative total energy with respect to a structure, the particle does not belong to the structure if it is located outside the structure's tidal radius. When a particle satisfies the membership conditions for multiple structures, we assign the particle to the least massive galaxies. The tidal boundary scheme is  implemented in \pgalf\ to separate close pairs of galaxies strongly interacting with each other. In a compact group of galaxies, for example, galaxies are likely to be strongly tied to each other with complex tidal structures, and the tidal boundary condition may help us to draw border lines between contacting galaxies. 

Table~\ref{tab:galaxy_number} shows the numbers of galaxies above a given stellar mass identified in the HR5 snapshots at $z=9-4$. Since stars form in high density gas cells located only in the zoomed region, galaxies are also identified in the region filled with grids of  level 13 or higher. However, low level/massive particles initially placed outside the zoomed region gradually permeate into the high resolution region due to the evolution of the density field as a function of cosmic time. In this study, we flag grids that contain low level DM particles ($M_{\rm DM}\ge 5.51\times10^8\,\msun$) and only use galaxies at least 3\,cMpc away from the contaminated regions, to minimize potential boundary effects caused by massive particles. One can find the details of the geometry evolution of the zoomed region in Appendix A of \citet{lee21}.

\begin{table}
    \centering
    \caption{Number of \hr\ galaxies  above given stellar mass limits. The numbers outside parentheses in the 2nd and 3rd columns are the counts of the galaxies locating at least 3\,cMpc away from low level particles (the clean set) while those inside are for the entire set. The fourth column present the volume uncontaminated by massive/low-level particles. In this study, we use the clean set only.}
    \begin{tabular}{crrc}
    \hline
    Redshift  & $M_{\star}>2\times10^9\,\msun$ &$M_{\star}>5\times10^{9}\,\msun$ & Volume \\ 
     &   &   & ($10^7$~cMpc$^3$) \\ 

       \hline
    9 & 14 (15) & - (-) & 1.3508\\
    8 & 71 (83) & 5 (6) & 1.3490 \\
     7 & 495 (593) & 58 (67) & 1.3459\\
     6 & 2205 (2669) & 1372 (446) & 1.3425\\
     5 & 8534 (10446) & 1918 (2348) & 1.3375\\
     4 & 22136 (27441) & 6095 (7590) & 1.3317\\

    \hline
    \end{tabular}
    \label{tab:galaxy_number}
\end{table}

\begin{figure*}
\centering 
\includegraphics[width=0.8\textwidth]{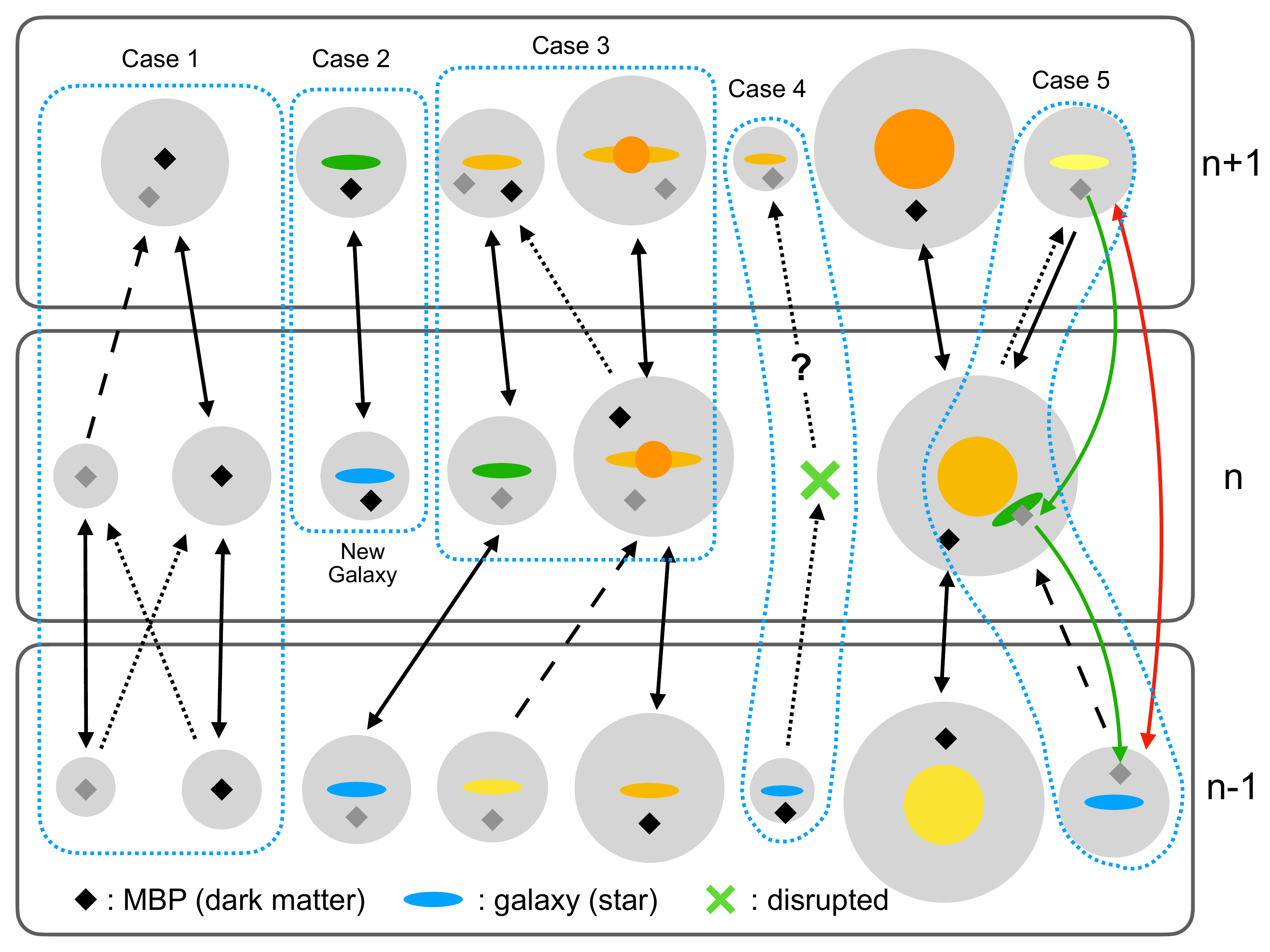}
\caption{Schematic diagram for the varying branches of merger trees constructed between snapshots $n-1$, $n$, and $n+1$. The grey shades are self-bound structures, the diamonds present the most bound particles (MBPs) in the self-bound structures and the colored ellipses are galaxies in them. The black solid arrows indicate main progenitor/descendant relations, dotted arrows mark the mass transfers between the structures with no main progenitor/descendant relations, and the dashed lines indicate mergers along the main descendant branches. The green solid arrows show the back-tracking of the MBPs to repair the branches broken by structure misidentification and the red solid arrow marks the main progenitor-descendant branch repaired by the MBP back-tracking scheme.}
\label{fig:merger_scheme}
\end{figure*}

\subsection{Merger Tree}

We construct the merger trees of self-bound objects (SBOs) throughout the entire snapshots of \hr\ based on the most-bound particles (MBPs) and stellar particles. The MBP is defined as the dark matter particle that has the least total (kinetic+potential) energy in a SBO~\citep{hong16}. We assume that the evolutionary history of a SBO can be traced by using the MBP, which, by definition, is most strongly bound to the host SBO. Since dark matter particles do not newly form or disappear in {\hr}, we use dark matter particles to identify the MBPs of SBOs.

Figure~\ref{fig:merger_scheme} shows schematic diagrams of some examples of various merger trees connecting SBOs in a series of three snapshots. Case 1 in this figure illustrates how the MBP scheme works for SBOs without stellar particles. The tree-building algorithm first identifies the MBP of each SBO at an initial time step, and then finds the SBO that possesses the interested MBP at the next time step. We then assign the substructure containing the MBP of the SBO at the next time step as the main descendant of the SBO. An SBO can acquire multiple MBPs from more than one SBOs in the earlier snapshot. In that case, we define the MBP donor with which the SBO shares its largest dark matter mass fraction as the main progenitor and classify the rest MBP donors as merged ones. Thus, in this algorithm, more than one SBOs at a given time step can have a common descendant in the next snapshot. The multiple MBPs coming from merged structures are tracked as the sub-MBPs along the main branches of their host SBOs throughout the entire snapshots. The main MBPs of SBOs are updated at every time step, because the internal kinematics of structures continuously changes. The MBP scheme enables us to trace the trajectories of SBOs that might have missed in-between the time steps due to structure mis-identifications that frequently occurs in dense environments~\citep{srisawat13}.

In \hr, we define a lump of stellar particles in a SBO as a galaxy. In principle, the MBP and the galaxy of an SBO should be closely bound to each other, but, in practice, they are sometimes decoupled due to complicated internal interactions, branching into different structures in the next step (See Case 3 in Figure~\ref{fig:merger_scheme}). In merger trees, this case can be seen, as if galaxies suddenly appear or disappear with no legitimate descendants or progenitors. To minimize such cases, we additionally use a tree building code \ysamtm~\citep{jung14,lee14} that forms merger tree branches based on individual particle IDs. \ysamtm\ has been originally developed to construct halo merger trees using dark matter particles that do not newly form or disappear throughout the duration of cosmological simulations. We have updated \ysamtm\ to be able to deal with stellar particles that also have unique IDs but newly formed in the simulation. \ysamtm\ first maps galaxies from a snapshot to the galaxies in the next snapshot that share at least one stellar particle. We define the main descendant/progenitor of a galaxy in a snapshot as the galaxy with which the  galaxy of interest shares its largest stellar mass fraction in the next/previous snapshot. If two galaxies in two adjacent snapshots are the main progenitor and descendant to each other, then the branch between them is classified as the main branch. A galaxy is regarded as merging into its main descendant when the main descendant has as the main progenitor another galaxy than the  galaxy of interest.

There are three cases that need to be addressed to clarify the evolutionary tracks of galaxies. i) A galaxy is classified as new when all its stellar particles form between the current and previous time steps, and it has no progenitors (Case 2 in Figure~\ref{fig:merger_scheme}). ii) Some galaxies have stellar particles that are not bound to any galaxy in the previous snapshot  but had already formed (Case 4 in Figure~\ref{fig:merger_scheme}). These types of galaxies are classified as originating from accretion. They are mostly contained in SBOs that have particle numbers to close to the minimum required to form FoF halos in our structure finding scheme. Therefore, such galaxies make a practically negligible contribution at galaxy masses above the completion limit of $M_{\star}\sim10^9\,\msun$ in \hr~\citep[see][]{lee21}. iii) A galaxy can have a main progenitor that connects with a main descendant which differs from the galaxy of interest (Case 5 in Figure~\ref{fig:merger_scheme}). Since the galaxies cannot have legitimate progenitors in this case, they are inevitably classified as new galaxies. This frequently occurs in fly-by galaxies in dense environments because their dynamical states are deeply entangled with central galaxies. Most of them soon escape, and are subsequently identified as independent SBOs. To patch the broken branches, we back-track the MBPs of the re-identified galaxies until the galaxies are explicitly separated from other SBOs in the past, as illustrated by the green arrows in Figure~\ref{fig:merger_scheme}. In this case, the legitimate progenitors are initially identified as those merging into other galaxies. The merger branches of such progenitors are revised, and connected to their legitimate descendants, instead of their massive neighbours%, based on the branches traced by their MBPs.

\begin{figure*}
\centering 
\includegraphics[width=0.85\textwidth]{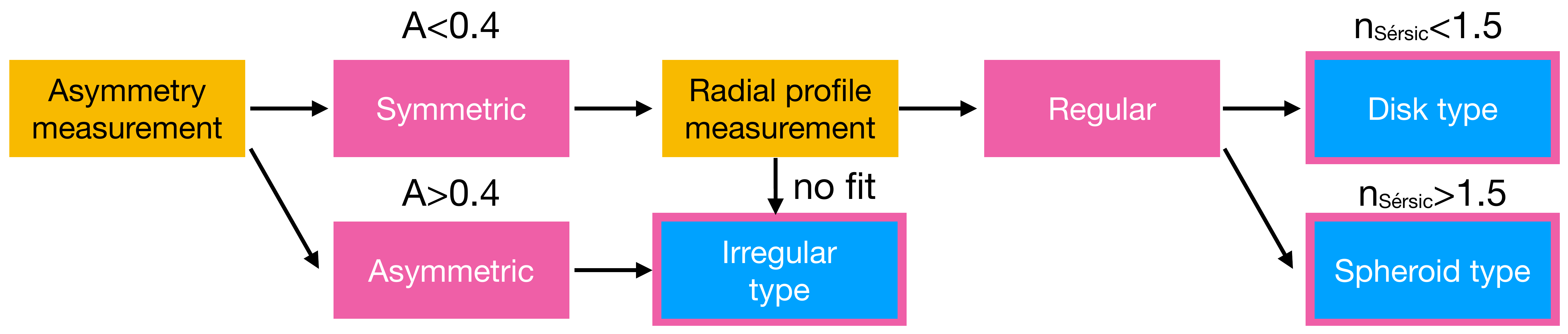}
\caption{Flowchart showing the morphology classification scheme. We first classify galaxies into symmetric and asymmetric ones using the asymmetry parameter and divide the symmetric galaxies into disks, spheroids, and irregulars based on their radial stellar-mass density profile parameterized by the S\'{e}rsic index $n_{\rm  S\acute{e}rsic}$. Galaxies that are not fitted to the S\'{e}rsic profile are classified as irregulars together with the asymmetric ones.}
\label{fig:scheme}
\end{figure*}

The merger trees are constructed based on the MBP scheme first, and branches are revised based on stellar particles if galaxies are present in the structures. Thus, the branches formed based on stellar particles are prior to those made based on the MBP, except for Case 5 in Figure~\ref{fig:merger_scheme}.

\section{Morphology of Galaxies}
We measure the galaxy morphology using  the asymmetry and the radial profile of their stellar-mass-density distribution. Therefore, the morphology we study here may be different from the morphology defined in observations which rely on the light distribution and take into account the locality of star formation within galaxies. As pointed out by \citet{trayford19}, albeit not a direct observable, the stellar-mass-density distribution provides us with the physical properties of galaxies, which must serve as a proxy to understand the origin and evolution of galaxy morphology. 
We assign morphological types to galaxies using the shape characteristics only, and avoiding other indirect information such as kinematics~\citep[as done by e.g.][]{Dubois2016, trayford19} or star formation activities. This is because the latter are not morphological properties, even though they can acquire some correlation with morphology in certain redshift ranges. Our morphology classification scheme is summarized in Figure~\ref{fig:scheme}.

We use the asymmetry parameter to classify galaxies into symmetric and asymmetric types. Radial stellar mass density profile is then used to measure the S\'{e}rsic Index, according to which symmetric galaxies are divided into disks and spheroids. Irregulars include asymmetric galaxies and those for which the S\'{e}rsic model fit fails. We do not use the axis ratios as shape parameters because they usually depend on distance from the center, are sensitive to shot noise in the outskirts, and thus are not robust.

\subsection{Asymmetry Parameter}
As shown in Figure~\ref{fig:scheme}, we first calculate the asymmetry parameter $A$ for each galaxy to decide if the galaxy has a symmetric or asymmetric morphology. We measure $A$ by adopting a methodology similar to that proposed by \citet{conselice00}. Specifically, we use the stellar-mass density of the galaxy defined on a three-dimensional mesh covering four times the root-mean-squared size of the galaxy. The stellar-mass density $\rho_{\star,i}$ at pixel $i$ is calculated by using a Spline kernel with a variable size containing 100 adjacent stellar particles. Then, we measure the asymmetry parameter from
\begin{equation}
 A = \frac {\sum_i w_i |\rho_{\star,i} - \rho_{\star,j}|}{2 \sum_i w_i \rho_{\star,i}},
 \end{equation}
where the pixel index $i$ goes through all density pixels in the mesh.
Here, pixels $i$ and $j$ are points reflected in the galactic center, and the weight $w_i$ is 1 when at least one of the pixels $i$ and $j$ contain more than 10 stellar particles and is 0 otherwise.
We also exclude the central region with a diameter of 1.6 pkpc, as this region is not well-resolved in \hr. 
The top left panels of Figures~\ref{fig:z7_prop} and \ref{fig:z5_prop} show the asymmetry parameter $A$ as a function of stellar mass $M_{\star}$, for all the galaxies identified at $z=7$ and 5, respectively. The galaxies with relatively high stellar mass seem to show a dichotomy in $A$ at around $A = 0.4 - 0.5$. Based on this finding,
%and following \citet{trayford19}, 
we classify galaxies with $A \ge 0.4$ as asymmetric galaxies and the rest as symmetric ones. 

Numerical smoothing of the density profile over the simulation resolution scale is not expected to affect the asymmetry parameter much, as we find that galaxies become irregulars when they actually undergo structure-destructing events such as close encounters or mergers. Therefore, the fraction of irregulars is not much expected to be biased.

\subsection{S\'{e}rsic Index}

We calculate the S\'{e}rsic index, $n_{\rm S\acute{e}rsic}$, defined as 
\begin{equation}
 \Sigma(R) = \Sigma_0 {\rm exp} (-k R^{1/n_{\rm S\acute{e}rsic}}), 
 \label{eq:defsersic}
\end{equation}
where $\Sigma (R)$ is the radial profile of the stellar-mass density projected on the galactic plane defined by the major and intermediate axes, and $R$ is the radius in the galactic plane~\citep{sersic63}.
%the major and intermediate axes plane of the galaxy, and $R$ is the radius in the corresponding cylindrical coordinate system~\citep{sersic63}. 
We find the axes by diagonalizing the moment of inertia tensor of the stellar system. We fit the radial-density profile of each symmetric galaxy having $A < 0.4$ with the function in Equation~\ref{eq:defsersic} to find the best-fitting $n_{\rm S\acute{e}rsic}$. 
The fit is made over the radial interval from 0.8 pkpc to the radius $R_{90}$ where 90\% of stellar particles are contained. Just like measuring $A$, we exclude the same central part to avoid spurious effects from the finite resolution of \hr~ and also to exclude influence of the galactic core, if any.
For some galaxies, $n_{\rm S\acute{e}rsic}$ is undefined (that is, fitting fails to converge) because their radial profile is inconsistent with a S\'{e}rsic model. They are then added to the irregular galaxy category along with asymmetric galaxies ($A>0.4$). Among the regular galaxies, those with $n_{\rm S\acute{e}rsic} > 1.5$ are called `spheroids', and those with $n_{\rm S\acute{e}rsic} < 1.5$ are called `disks'.

This choice is made based upon the S\'{e}rsic index of observed galaxies. We have measured the S\'{e}rsic indices of the SDSS galaxies in the training set of \citet{park05}, and find that $n_{\rm S\acute{e}rsic} = 1.5$ divides the galaxies  into early and late types with a high reliability.
Figure~\ref{fig:iso_sersic} in Appendix~\ref{sec:sersic_index} shows the distribution of each type of the SDSS galaxies in the isophotal axis ratio versus the S\'{e}rsic index plane.

$R_{90}$ containing 90\% of total galaxy stellar mass is the outer limit of the radial range where the mass density profile is calculated and used to estimate the S\'{e}rsic Index.
The medians of $R_{90}$ are 2.8, 3.2, 3.6, and 4.4 pkpc at $z=$7, 6, 5, and 4 for our least massive galaxies with $M_{\star}\sim2-4\times10^9\,\msun$, respectively. The 10\% of stellar particles at the outermost region are dropped to minimize the effects of mass contribution of possible stray galaxies and tidal perturbation of neighboring galaxies.
On the other hand, the inner limit is set to $r=0.8$~pkpc taking into account the force resolution of \hr.   The finite simulation resolution is expected to make galaxies less concentrated at their center. Its impact on morphology classification is mitigated because the central part is excluded. 
The radial interval for density profile calculation is significantly larger than the simulation resolution scale for relatively more massive galaxies with $M_{\star} \ge 5\times 10^9 M_{\odot}$ and at lower redshifts $z\leq 6$.

\begin{figure}
\centering 
\includegraphics[width=0.45\textwidth]{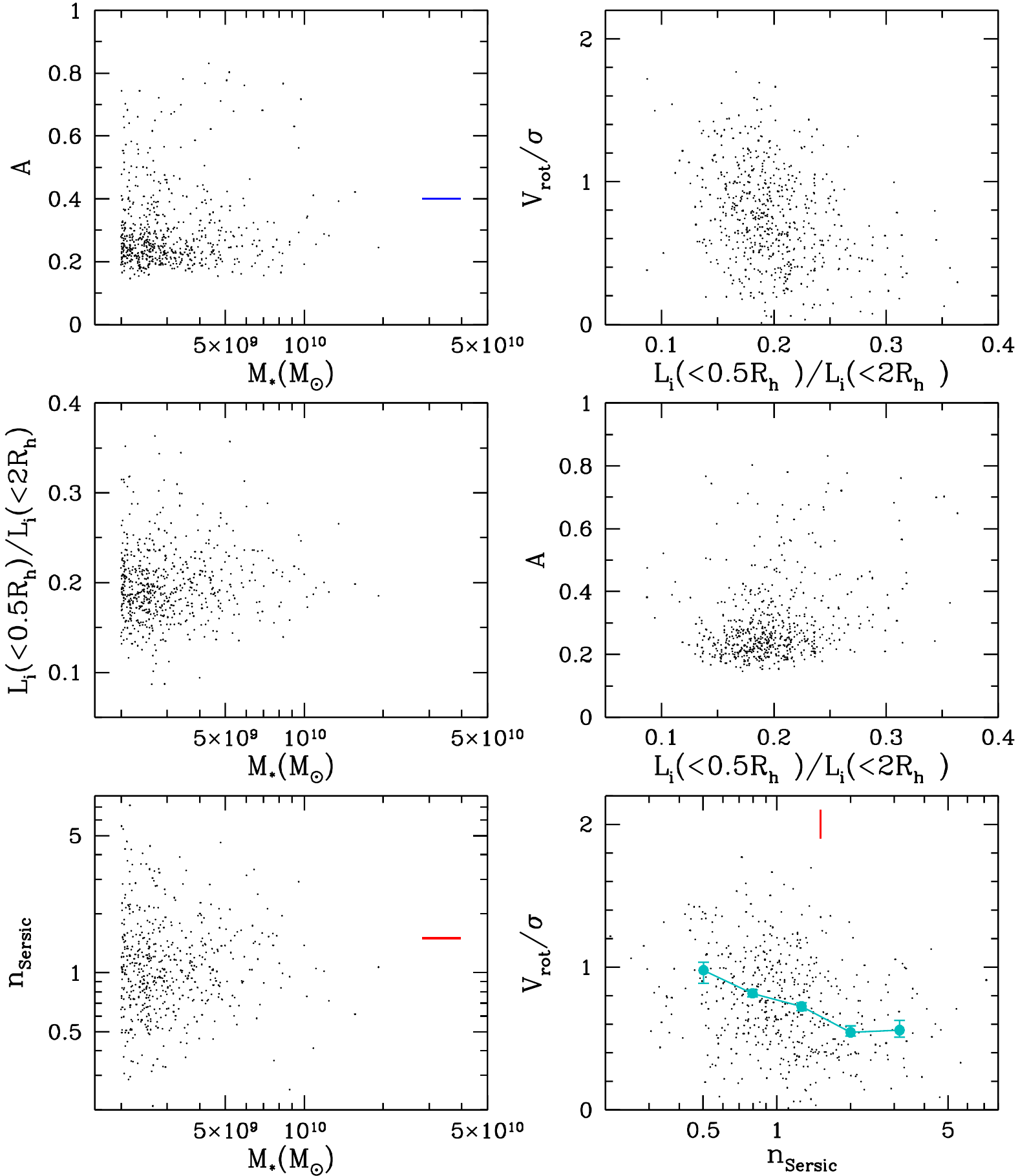}
\caption{Relations between the asymmetry parameter $A$,  $i-$band concentration parameter, S\'{e}rsic index $n_{\rm S\acute{e}rsic}$, ratio of rotational velocity to velocity dispersion $V_{\rm rot}/\sigma$, and stellar mass $M_{\star}$ of galaxies identified at $z=7$. The short blue line in the top left panel marks the asymmetry parameter cut of $A=0.4$ dividing symmetric and asymmetric galaxies. The short red lines in the bottom panels denote the S\'{e}rsic index cut demarcating disk-type and spheroid-type galaxies. The cyan curve in the bottom right panel shows the median $V_{\rm rot}/\sigma$ at given $n_{\rm S\acute{e}rsic}$, and error bars are the uncertainty limits of the medians.}
\label{fig:z7_prop}
\end{figure}

\begin{figure}
\centering 
\includegraphics[width=0.45\textwidth]{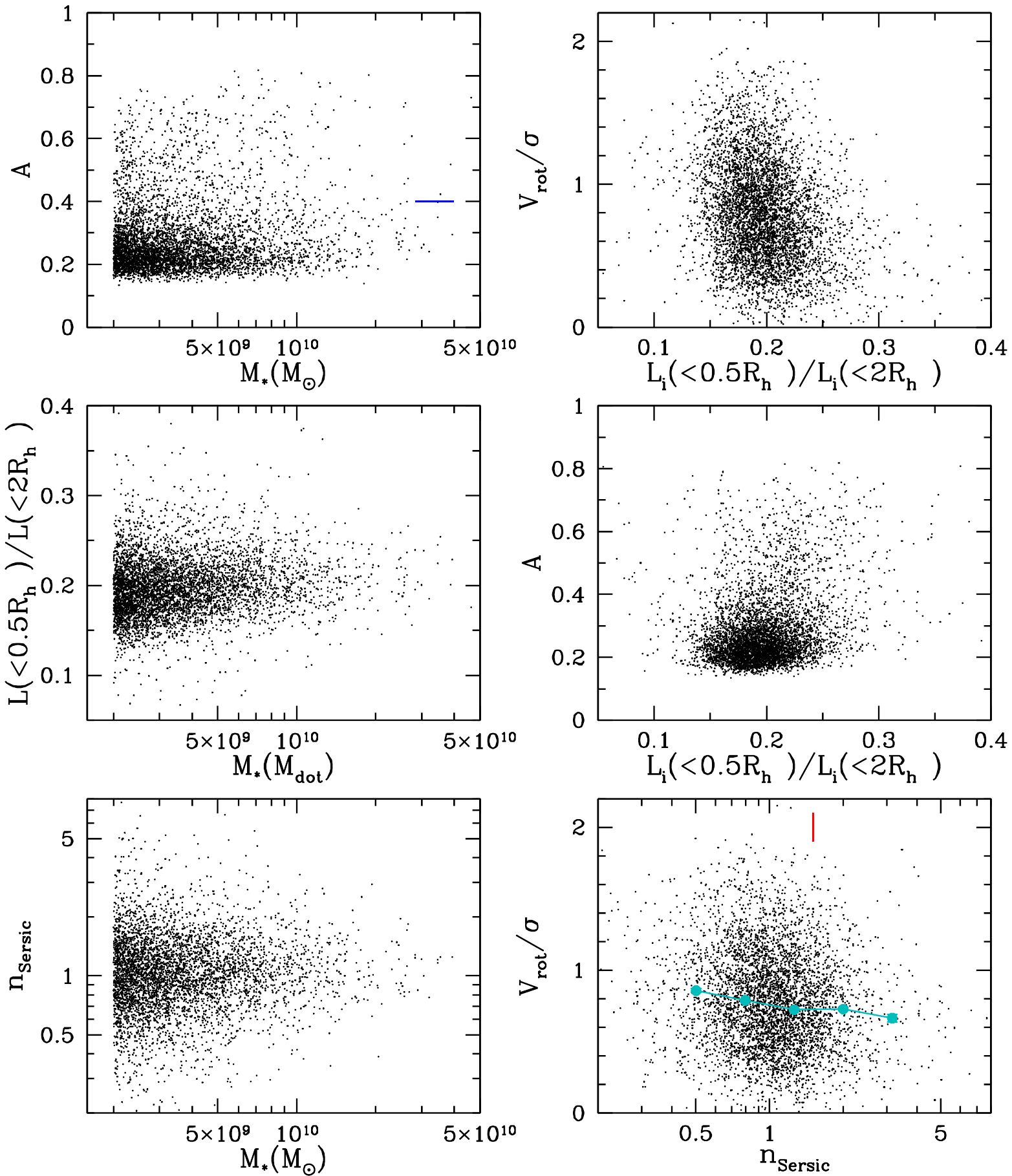}
\caption{Same as in Figure~\ref{fig:z7_prop}, but for galaxies at $z=5$. Only half the sample in $x<0.5L_{\rm box}$, where $L_{\rm box}$ is the side length of the \hr\ volume, is plotted here to make the figure less saturated.}
\label{fig:z5_prop}
\end{figure}

\begin{figure}
\centering 
\includegraphics[width=0.47\textwidth]{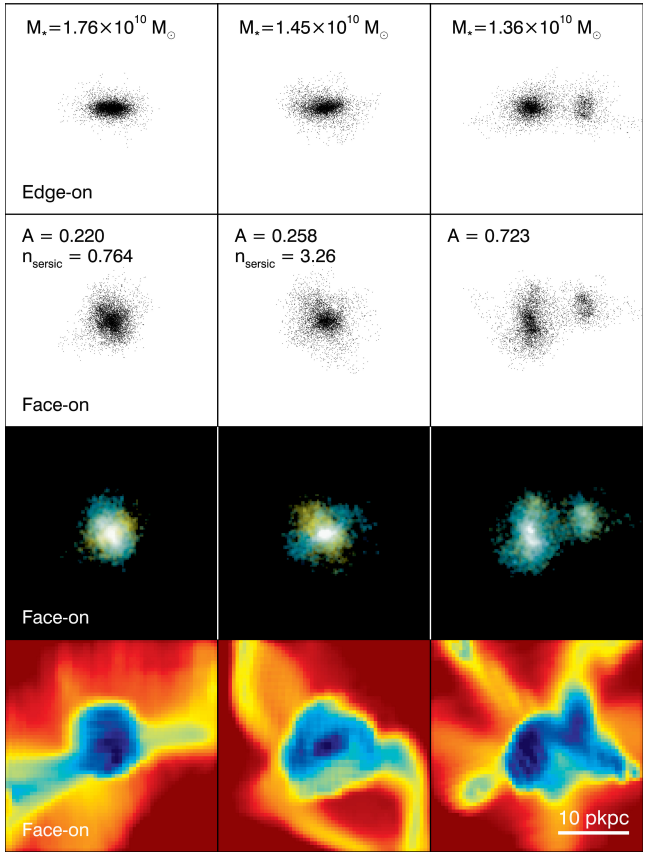}
\caption{Distributions of stars and gas of three galaxies identified at $z=6$. First and second rows show edge-on and face-on views of stellar particle distributions, respectively, the third row shows stellar mass density distribution colored according to their ages (bluer is younger), and bottom panels show gas density distribution. Gas density increases from red to blue color. The scale bar is in units of proper kpc. $M_{\star}$, $A$, and $n_{\rm S\acute{e}rsic}$ are stellar mass, asymmetry parameter, and S\'{e}rsic index, respectively.}
%\caption{Face-on views of three galaxies identified at $z=6$. Left panels show stellar particles, the middle column shows stellar mass density distribution colored according to their ages (bluer is younger), and right panels show gas density distribution. The gas density increases from red to blue color. The scale bar is in units of proper kpc. Asymmetry parameter $A$, and S\'{e}rsic index $n_{\rm S\acute{e}rsic}$, and stellar mass $M_{\star}$ are given in the left column.}
\label{fig:face_on}
\end{figure}

The bottom panels of Figures~\ref{fig:z7_prop} and \ref{fig:z5_prop} show the S\'{e}rsic index as a function of stellar mass (left panel) and the rotational velocity to velocity dispersion ratio ($V_{\rm rot}/\sigma$) as a function of S\'{e}rsic index (right panel), at $z=7$ and 5, respectively. The velocity components of stellar particles are measured in a cylindrical coordinate system in which the direction of the angular momentum vector of stellar mass is chosen to be the $z$-axis. The rotational velocity $V_{\rm rot}$ is the mean tangential velocity of stellar particles inside the cylindrical shell of an inner radius $R_{1/3}$ and an outer radius $R_{2/3}$, where $R_{1/3}$ and $R_{2/3}$ are the radii containing one and two third the total stellar mass measured from the central axis, respectively. The velocity dispersion $\sigma$ is defined as $\sigma=\sqrt{(\sigma^2_R+\sigma^2_{\rm rot}+\sigma^2_z)/3}$, where $\sigma_R$, $\sigma_{\rm rot}$ and $\sigma_z$ are velocity dispersion in the radial, tangential, and vertical directions.
The bottom right panel of Figure~\ref{fig:z7_prop} shows that galaxies with large $n_{\rm S\acute{e}rsic}$ have smaller $V_{\rm rot}/\sigma$ at $z=7$, but the relationship becomes weaker at $z=5$ in the bottom right panel of Figure~\ref{fig:z5_prop}.  It is interesting to see that the relation is tighter at high redshifts and then nearly disappears at $z < 5$ for these relatively small mass galaxies. 
%We will see in Discussion section that the first galaxies with $M_{\star} > 2\times 10^9\,\msun$ start to merge each other along filaments after $z\sim 6$.
This may be because of various evolutionary effects such as galaxy merger, as examined in the Discussion section.
Figure \ref{fig:z7_prop} and \ref{fig:z5_prop} also show the relationships between various physical parameters including the $i$-band concentration index $L_i(<0.5 R_{\rm h})/L_i(<2R_{\rm h})$, where $L_i(<R)$ is the total $i-$band luminosity within a radius $R$ and $R_{\rm h}$ is the half-light radius in a wavelength band.
%the ratio of the luminosities within $0.5 R_{\rm h}$ and $2R_{\rm h}$, where $R_{\rm h}$ is the half-light radius. 

Figure~\ref{fig:face_on} shows edge-on and face-on views of representative disk (left), spheroid (middle), and irregular (right) type galaxies at $z=6$. The irregular galaxy is undergoing a merger, and has a large asymmetry parameter $A=0.723$.
The spheroid galaxy in the middle panels ($n_{\rm S\acute{e}rsic}=3.26$) shows a much more compact core than that of the disk galaxy ($n_{\rm S\acute{e}rsic}=0.764$) whose stellar-mass density is extended. Figure~\ref{fig:face_on} demonstrates that we are able to conduct morphology classification of galaxies identified in the \hr\ simulation following the scheme shown in Figure~\ref{fig:scheme}.

In this section we have shown how the asymmetry parameter and S\'{e}rsic index are related with other structural and kinematic parameters. Figures 3 and 4 show that the majority of galaxies are symmetric ones with $A$ between 0.15 and 0.4, with an extended tail of asymmetric galaxies above $A\approx 0.4$. $n_{\rm S\acute{e}rsic}$ has a continuous distribution across 1.5 and is nearly independent of $M_{\star}$ up to $\sim 5\times 10^{10}\,\msun$. $V_{\rm rot}/\sigma$ is a decreasing function of the concentration parameter or $n_{\rm S\acute{e}rsic}$ at high redshifts, which weakens at later epochs.

\begin{figure}
\centering 
\includegraphics[width=0.45\textwidth]{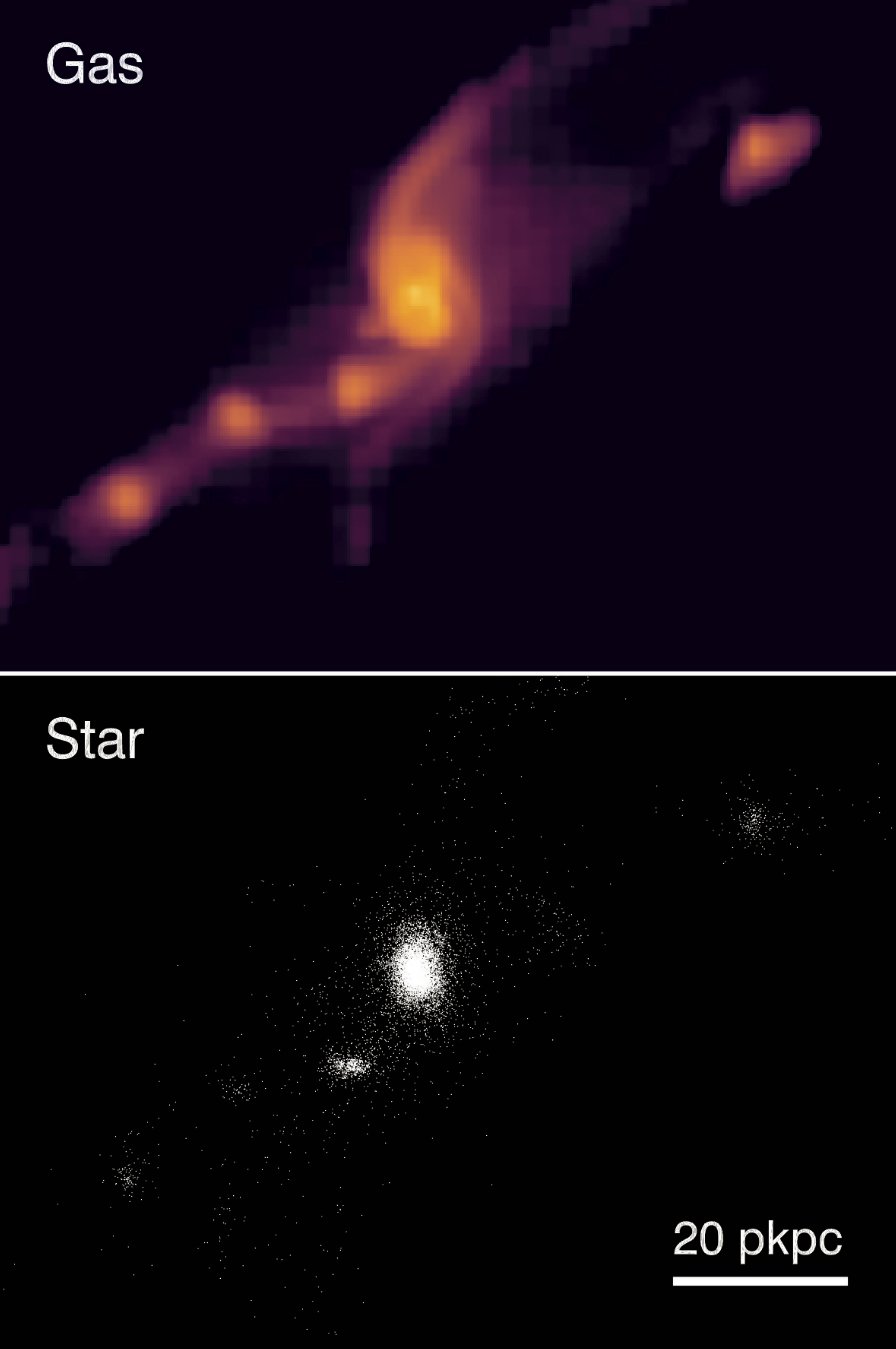}
\caption{Gas density (top) and stellar particle distribution of a small group at $z=6$. The group of galaxies are connected to each other via a gaseous filament. The central galaxy has stellar mass of $M_{\star}=1.76\times10^{10}\msun$, and its gas shows spiral arms. Small objects in the neighbour have stellar mass smaller than $10^9\msun$.} %\st{The group of galaxies are connected to each other via a gaseous filament.} 

\label{fig:galaxy_map}
\end{figure}

As shown below, the first galaxies in the cosmic morning are dominantly disks. In addition, we find that there are galaxies having spiral arms at this epoch. For these galaxies, the spiral arms are weak in the stellar-mass distribution, but are often quite prominent in the gas-density distribution. An example is shown in Figure~\ref{fig:galaxy_map}.
It is interesting to note that the morphological characteristics of the present-day spiral galaxies, i.e., spiral arms or disks, have counterparts in the early cosmic history, even though the mechanism for their origin may not be the same~\citep{Dubois2016, park19}.

\begin{figure*}
\centering 
\includegraphics[width=1\textwidth]{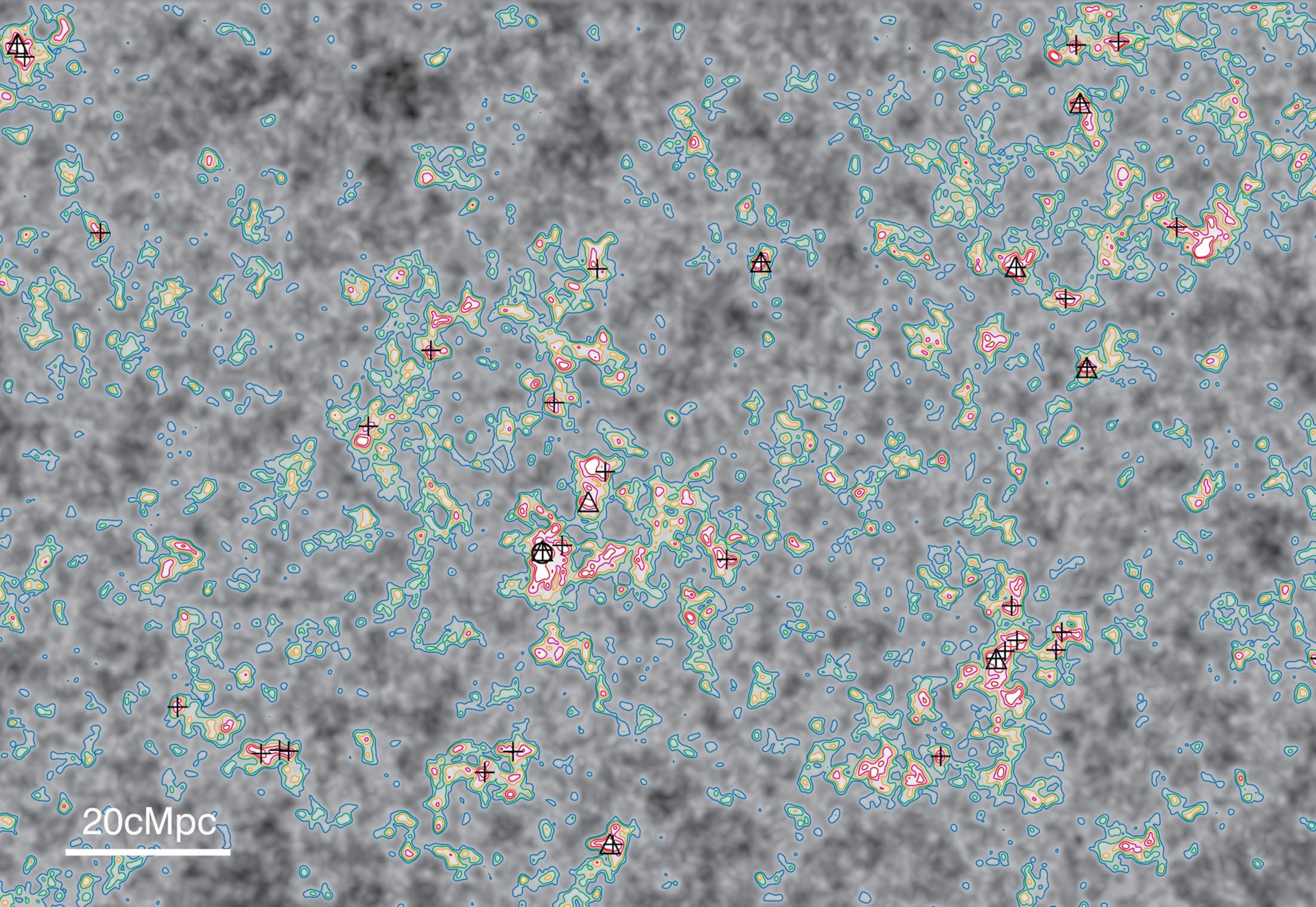}
\caption{A 1.66-cMpc-thick slice of \hr's initial dark-matter density field.
{\it Gray scale map}: The highest density within the slice along the direction perpendicular to the slice, after smoothed over 0.35 cMpc with the Gaussian filter.
{\it Color lines}: Contour levels correspond to the overdensity in units of the RMS overdensity, $\nu_{\rm DM}\equiv(\rho_{\rm DM}/\bar{\rho}_{\rm DM}-1)/\sigma_{\rm DM}=$ 1.5 (blue), 2.0 (green), 2.5 (orange), 3.0 (magenta), and 3.5 (red).
{\it Symbols}: The formation sites of the first galaxies, defined as the initial center of mass of dark-matter particles that form the dark-matter halos of galaxies identified at $z=7$ (the circle), 6 (triangles), and 5 (crosses). The large-scale structures in the universe emerge at the highest density regions and grow in time as galaxy formation sites move from high to relatively lower density peaks.}
\label{fig:density_field}
\end{figure*}

\section{Results}
\subsection{Sites of Galaxy Formation in the Cosmic Morning}

By tracing the dark-matter particles belonging to galaxies back to their initial position, we find that almost all high-redshift galaxies form at high density peaks in the initial density field. We also find that the galaxy-formation site in the initial density field migrates starting from the highest density peaks to relatively lower density peaks.
%This is contrary to \citet{porciani02}'s report that the association between the centers of proto-halos and the initial density is vague and only about 40\% of proto-halos contain peaks in the initial density field. 
%\Xtophe{ Did Portiani refer to high redshift galaxies:  at high mass he did find a good match, so the contradiction might be less stringent? } 
\begin{figure*}
\centering 
\includegraphics[width=0.69\textwidth]{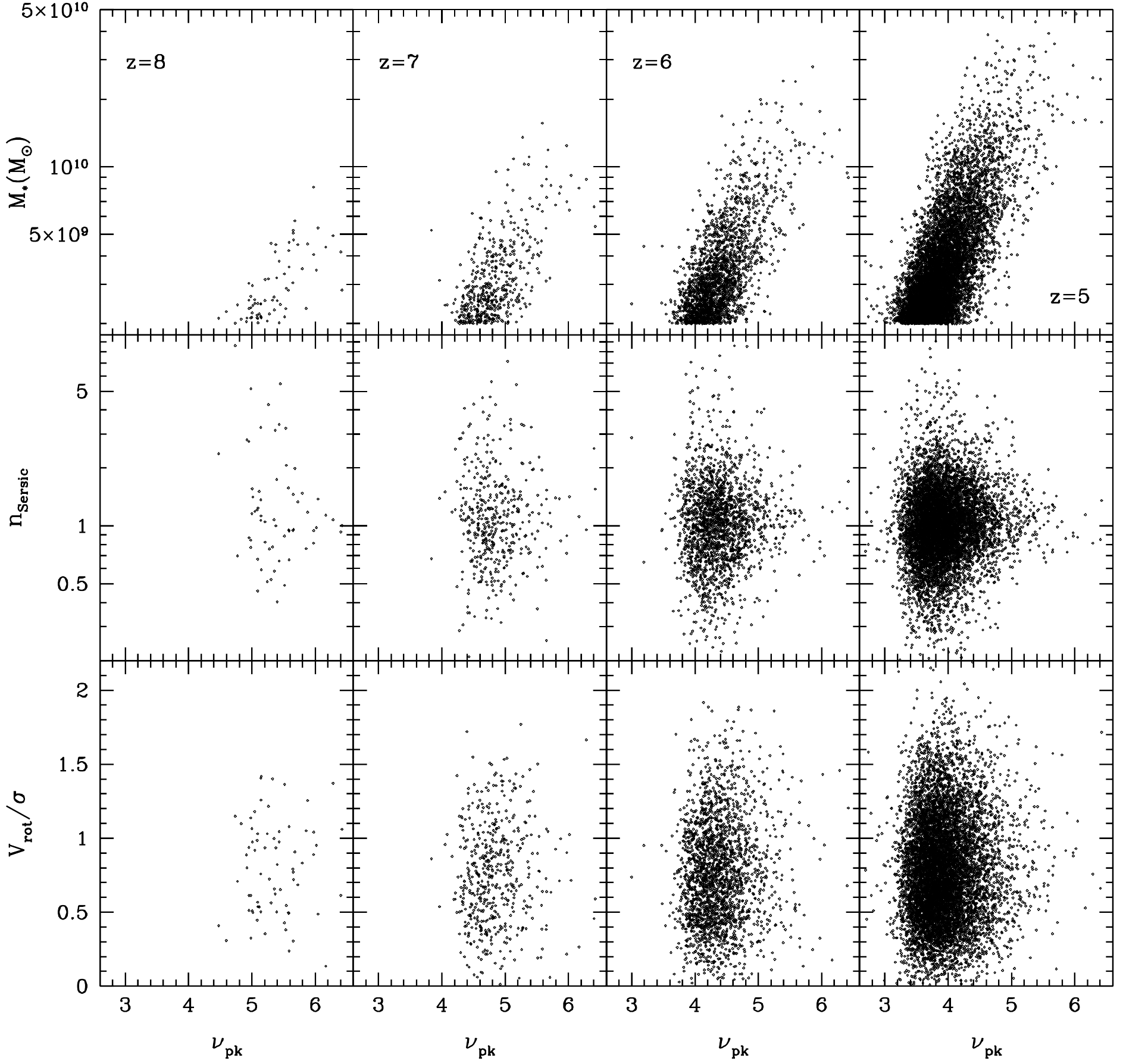}
\caption{Stellar mass (top), S\'{e}rsic index (middle), and rotational velocity to velocity dispersion ratio (bottom) of galaxies at redshifts 8, 7, 6, and 5 as a function of the initial dark matter overdensity peak height $\nu_{\rm pk}=(\rho_{\rm DM,pk} - {\bar \rho_{\rm DM}})/\sigma_{\rm DM}$ matched to the initial center of mass of the dark matter in galaxies at each epoch. }
\label{fig:prop_dm_density}
\end{figure*}
%
%We find that the observed stellar cosmic web  today is a snapshot of the large-scale structures in the universe only at the present epoch.

\subsubsection{Galaxy-formation sites in initial conditions}

To find out where in the initial conditions galaxies would form later at a redshift $z$, we use the dark-matter particles belonging to the galaxies identified at that redshift. 
%We identify galaxies and the dark matter particles associated with them at that redshift. 
Specifically, we find the initial positions of these dark-matter particles and define the initial center, that is, the galaxy-formation site, for each galaxy by the center of mass of the dark-matter particles' initial positions.

We then match the initial galaxy-formation sites to the peaks in the initial density field as follows. First, we smooth the initial density field with a Gaussian filter over the smoothing length $R_G=0.35$ cMpc, which is 2.73 times the density pixel size. Then density peaks are identified. Starting from the most massive galaxies, we search for the highest unmatched peak within $r = 4.5R_G=$ 1.575 cMpc ball around each initial galaxy-formation site. The search proceeds for the lighter galaxies after excluding the peaks already chosen.
%We then search for the nearest initial density peak starting from the most massive galaxies by finding the highest peak within $4.5 R_G=$ 1.575 cMpc, where $R_G = 0.35$ cMpc is the Gaussian smoothing length of the smoothed initial density field. This smoothing length is 2.73 times the density pixel size. The peaks already chosen for massive galaxies are excluded in the search for the peaks corresponding to less massive galaxies. 
With this method, we have successfully identified, for example, the 99.6\% of the galaxies with $M_{\star} > 2\times 10^9 \,\msun$ at $z=6$ to the initial density peaks.

Figure~\ref{fig:density_field} shows the initial dark-matter density field of \hr\ in a 1.66 cMpc-thick slice. Contour levels delineate the %significance 
dark matter overdensity threshold $\nu = (\rho_{\rm DM} /{\bar \rho}_{\rm DM} -1) / \sigma_{\rm DM}$ 
%of dark-matter density above the mean 
in units of 
%with respect to 
the standard deviation $\sigma_{\rm DM}$. Symbols are the initial center of mass of dark-matter particles that belong to the dark-matter halos of galaxies identified at $z=7$ (circle), $6$ (triangles), and $5$ (crosses). The overlap of all the three symbols on top of a peak with $\nu_{\rm pk}=5.34$ at the 7 o'clock from the center of Figure~\ref{fig:density_field} indicates that the galaxies identified at all three redshifts share the same initial density peak. 
%This is the same galaxy formed out of the noted initial density peak,  keeping its identity. 
We also find that the formation sites for the $z\ge7$ galaxies correspond to the locally highest peaks residing on globally over-dense regions. One typical example is shown in the Figure~\ref{fig:density_field} as a circled region. This must be contrasted to the formation sites for $z=5$ galaxies (crosses) that correspond to relatively lower initial peaks that are sitting on filaments connecting the higher over-density regions, showing the growth and emergence of large-scale structures of galaxies out of underlying evolving matter distribution.

\subsubsection{Initial peak height and galaxy properties}

The top panels of Figure~\ref{fig:prop_dm_density} show the relation between the stellar mass of the galaxies identified at $z=8$, $7$, $6$, and $5$ (from left to right), and the peak height of the corresponding initial formation sites. Two evolutionary features are clearly seen here: the migration of the formation sites toward low-density regions and the increase of galaxy stellar mass. It should be noted that at a particular epoch the galaxies with $M_\star \approx 2\times 10^9 M_{\odot}$ have originated from the initial density peaks with a narrow range (RMS $\Delta \nu_{\rm pk}=$ 0.23, 0.32, and 0.53 from the mean ${\bar \nu}_{\rm pk}=4.60, 4.11, 3.58$ at $z=7$, 6, and 5, respectively) in their initial peak height.

\begin{figure}
\centering 
\includegraphics[width=0.45\textwidth]{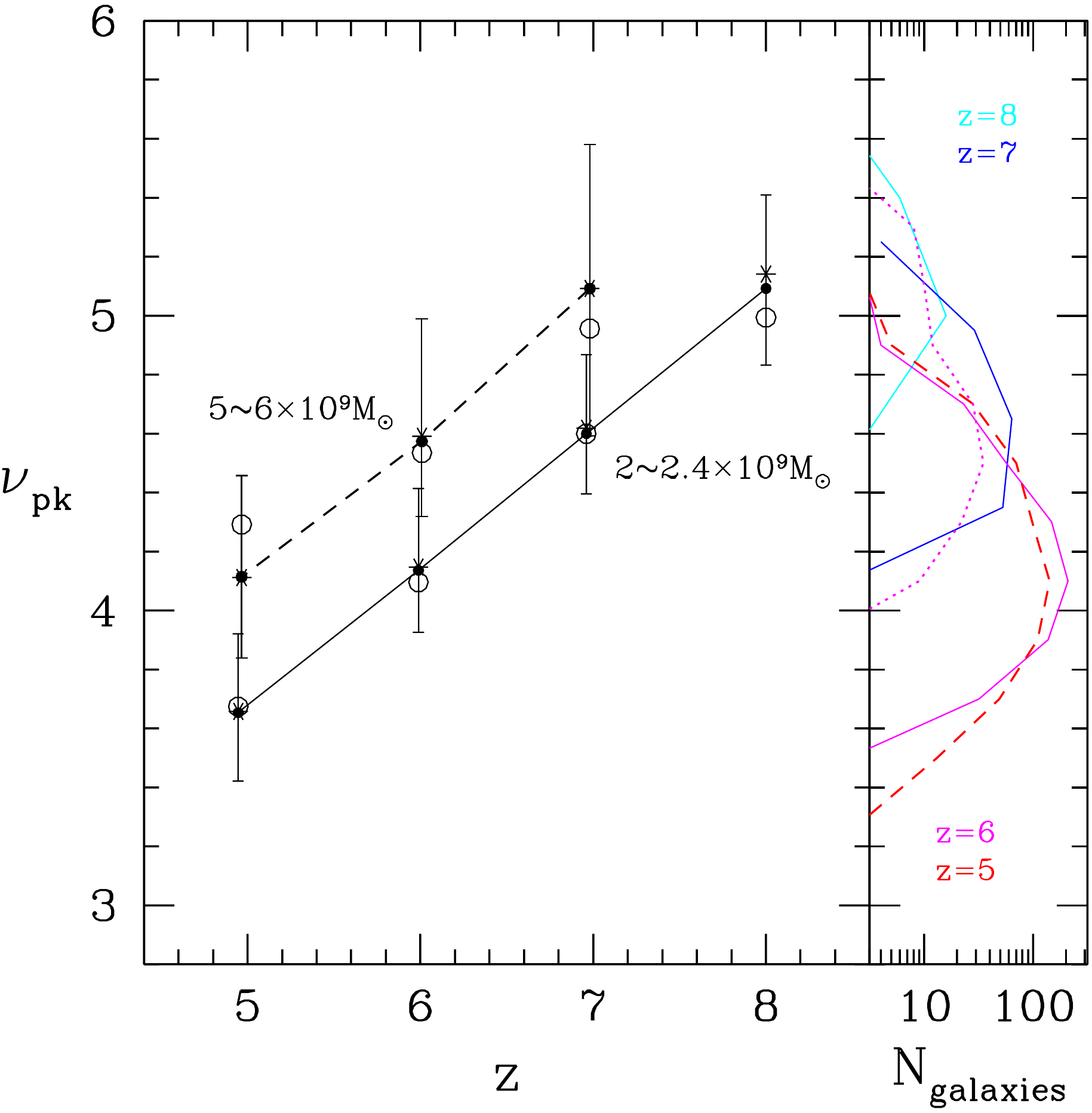}
\caption{Left: Initial dark matter overdensity peak height $\nu_{\rm pk}$ corresponding to the center of mass of the dark matter in the galaxies having a given stellar mass. The solid and dashed lines are for the galaxies with $M_{\star}=2-2.4\times10^9\,\msun$ and $M_{\star}=5-6\times10^9\,\msun$, respectively. The error bars show $16^{\rm th}-84^{\rm th}$ percentile distributions. The open circles mark the median $\nu_{\rm pk}$ of spheroids, and the asterisks are for disk-type galaxies. Right: Histograms of galaxy number count as a function of $\nu_{\rm pk}$. The cyan, blue, and magenta solid curves correspond to the galaxies with $M_{\star}=2-2.4\times10^9\,\msun$ at $z=8$, 7, and 6. The magenta dotted and red dashed curves show the distributions of the galaxies with  $M_{\star}=5-6\times10^9\,\msun$ at $z=6$ and 5, respectively.}
\label{fig:dm_density}
\end{figure}

To quantify the evolution, we show the mean and standard deviation of the peak heights as a function of redshifts for two narrow stellar-mass ranges in the left panel of Figure~\ref{fig:dm_density}. The plot shows that galaxies with a fixed stellar mass form at lower and lower initial density peaks as a redshift decreases. The galaxies having $M_{\star} = 2.0 - 2.4 \times 10^9\,\msun$ at an epoch $z$ have formed typically out of the initial density peaks with mean height ${\bar \nu}_{\rm pk} = 0.472 z +1.32 \pm 0.01$. Massive galaxies are just those that have formed earlier (at higher redshifts) with smaller mass, then acquired stellar mass through internal star formation, accretion or merger. The galaxies with $M_{\star} =5-6 \times 10^9 \,\msun$ at $z=5$, for example, are roughly those with $M_{\star} =2-2.4 \times 10^9 \,\msun$ at z=6, and have formed out of the initial density peaks with $\nu_{\rm pk} \approx 4.1$.

A close inspection of the distribution of initial peak heights supports the above scenario. The three solid curves in the right panel are the distributions of $\nu_{\rm pk}$ of the peaks corresponding to the galaxies with $M_{\star} =2-2.4 \times 10^9 \,\msun$ at $z=8$ (top), 7 (middle), and 6 (bottom). The bottom curve for $z=6$ is quite similar to the dashed curve, which is the distribution function of  galaxies with $M_{\star} =5-6 \times 10^9 \,\msun$ at $z=5$ supporting again the above observation. The initial density peaks with $\nu_{\rm pk} \approx 4.1$ collapse to form $M_{\star} =2-2.4 \times 10^9 \,\msun$ galaxies by $z=6$, and grow to reach $M_{\star} =5-6 \times 10^9 \,\msun$ without losing their identity until $z=5$. On the other hand, the distribution function for those with $M_{\star} =5-6 \times 10^9 \,\msun$ at $z=6$ (dotted curve) is somewhat skewed to high $\nu_{\rm pk}$ and is rather different from the middle solid curve, even though the mean peak heights are close ($\approx 4.6$).

The second and third rows of Figure~\ref{fig:prop_dm_density} show that the S\'{e}rsic index and $V_{\rm rot}/\sigma$ of galaxies do not clearly depend on the initial peak density. We will return to this question of the effects of the initial conditions on galaxy morphology and internal kinematics in section 4.4.

\begin{figure}
\centering 
\includegraphics[width=0.38\textwidth]{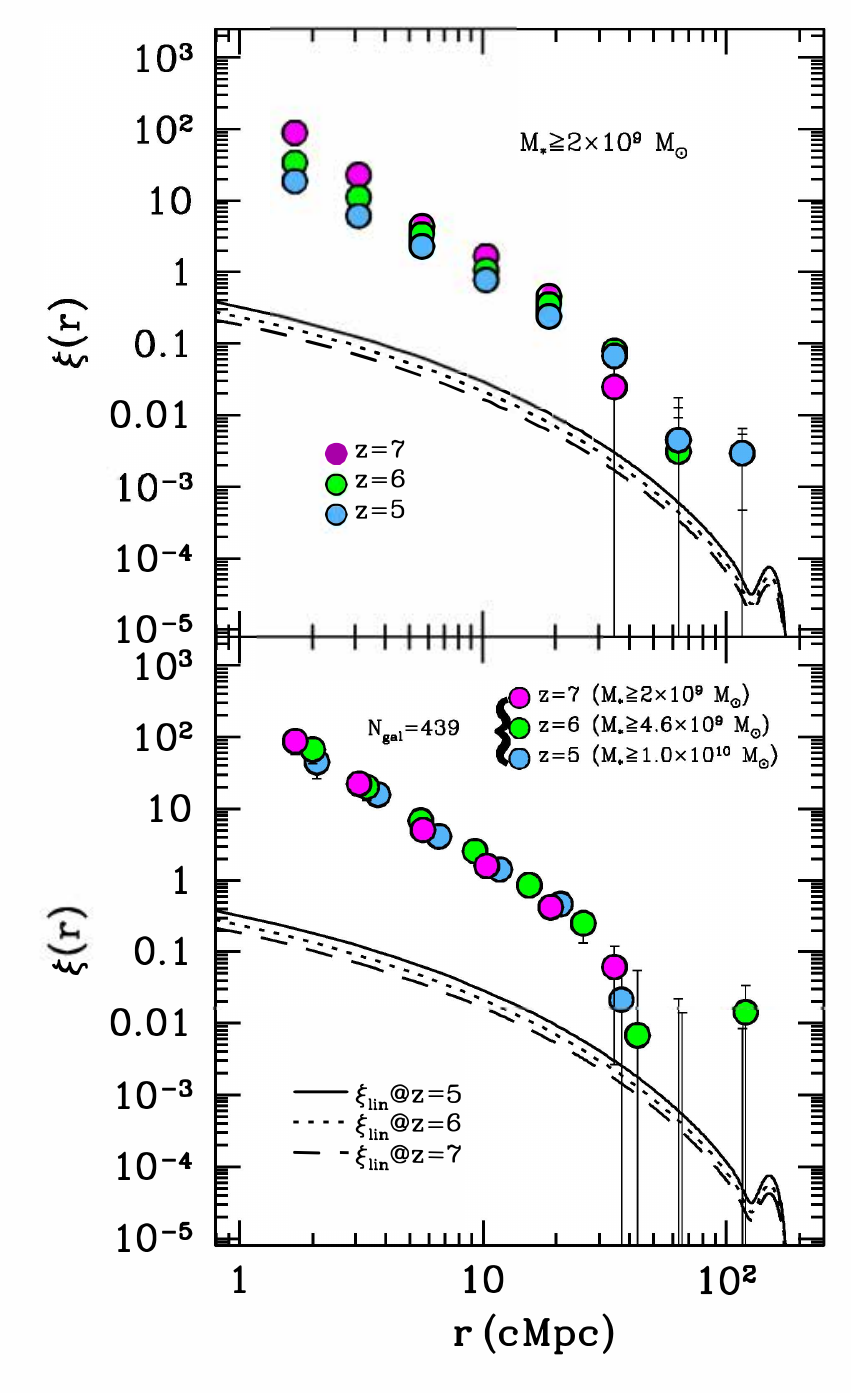}
\caption{{\it Top:} The two-point correlation function for the galaxies with $M_{\star} \geq 2\times 10^9\,\msun$ at three different redshifts, $z=7$ (Magenta), $6$ (Green), and $5$ (Blue). {\it Bottom:} the two-point correlation function at the same three redshifts but with different stellar-mass cuts, to sample the same number of the most massive galaxies, giving the fixed number of galaxies of $N_{\rm gal}=439$. For each panel, we also show the linear matter two-point correlation function at the same redshifts as solid line ($z=5$), dotted line ($z=6$) and dashed line ($z=7$).
The stronger evolution of the small-scale nonlinear bias shown for the fixed stellar-mass-limit case ({\rm Top}) has diminished when fixing the total number of galaxies.
} 
\label{fig:gal_CF}
\end{figure}

How is galaxy bias related with the peak height of the initial formation site? Figure \ref{fig:gal_CF} shows the galaxy two-point correlation functions at three different redshifts ($z=7$, $6$, $5$), for fixed stellar mass cut ($M_{\star}>2\times 10^{9}\,\msun$, top panel) and for a fixed total number of galaxies ($N_{\rm gal}=439$, bottom panel).
As shown in Figure~\ref{fig:dm_density}, again, for a fixed stellar-mass limit, the high-$z$ galaxies have formed at the initial density peaks with higher $\nu_{\rm pk}$, and the galaxies formed at higher $\nu_{\rm pk}$ (at the rarer peak) are more biased and strongly clustered. We can see this trend most clearly for $r<20\,{\rm cMpc}$ in the upper panel of Figure~\ref{fig:gal_CF}, where the clustering amplitude is larger for higher-redshift galaxies. On the other hand, when fixing the total number of galaxies, the significance $\nu_{\rm pk}$ %corresponding to the stellar-mass limit 
remains approximately the same. That is, the same galaxies maintaining their identity are sampled at the three redshifts. In this case we may approximate that the galaxy bias evolves passively~\citep{Fry:1996,Tegmark/Peebles:1998}, for which case the density contrast of highly-biased ($b\gg1$) galaxies stays constant. As a result, the two-point correlation function of these galaxies does not evolve in time, as shown in the bottom panel of Figure~\ref{fig:gal_CF}.

%The linear bias evolves as $b(z)=[b(z_*)-1]D(z_*)/D(z) + 1$, and, for as is the case in, we can approximate the passive evolution as 
%$b(z)\simeq b(z_*)D(z_*)/D(z)$. Then $\xi(r,z)\propto b(z)^2D^2(z)/D^2(z_*)\xi_m(r,z_*)$ stays constant in time. Indeed, the bottom panel of Figure~\ref{fig:gal_CF} shows that the clustering amplitude is independent of time on linear scales ($\xi\ll1$), and the trend continues for the nonlinear scales as well.

\subsubsection{Emergence of the first large-scale structures}

We note that the large-scale structures in the universe as traced by galaxies emerge in the cosmic morning in a way similar to
%Consequently, %the scenery of the world of galaxies has been evolving since cosmic morning in a way similar to 
%the scenery of galaxy evolution since cosmic morning is in a way similar to 
the process that rhizomes spread out roots %underground 
and shoot out culms from their nodes on the way. Burst of galaxy formation starts first in the highest density regions, the locations of the first {\it cosmic rhizomes}, in which the first galaxies appear at the rare highest peaks. As the underlying dark matter and gas collapse into filaments, the formation sites of the next new galaxies propagate away from the highest density regions along attached filaments, increasing the realm of each rhizome, see Figure~\ref{fig:density_field}. 

Galaxies at the outer boundary of rhizomes are in a different formation and evolution environment from the earlier ones and therefore may have different physical properties. The initial birth places of the first rhizomes are likely to evolve to protoclusters, as galaxies flow into them along filaments or as filaments collapse onto them. As time passes and matter fluctuations grow, initially lower density peaks also become non-linear and new less prominent rhizomes appear. And the first generation rhizomes become multiply connected and continue to spread out toward less dense regions. The full cosmic web of galaxies appears when most cosmic rhizomes percolate with one another and become globally connected~\citep[see also][]{gott86,shandarin89,park90,vogeley94,Bond1996}.
The picture depends strongly on the mass of galaxies used for defining large-scale structures. For example, the most massive galaxies live in disconnected regions even today and never form a web structure. 

The results that the initial $\nu_{pk}$ of newly forming galaxies decreases in time (Figures~\ref{fig:prop_dm_density} and \ref{fig:dm_density}), and monotonic decrease of the correlation function of galaxies with fixed stellar mass (Figure~\ref{fig:gal_CF}) quantitatively support the cosmic rhizome picture of the growth of large-scale structures in the universe. Such a description of cosmic-web formation is broadly consistent with the theories based upon the threshold density~\citep{Kaiser1984}, such as peak theory~\citep{BBKS1986} or excursion set theory~\citep{Bond1991}, although the detailed mathematical modeling requires
further understanding of galaxy-formation physics~\citep[following, e.g.][or more generally a formulation in terms of redshift dependant galactic bias]{Musso2018}.
Yet, it does not fully capture the dynamical top-down scale coupling for the gas, which  funnels galaxy morphology 
through cold streams~\citep{Pichon2011}, as discussed below.

\subsection{Morphological Types of Galaxies}

\begin{figure}
\centering 
\includegraphics[width=0.45\textwidth]{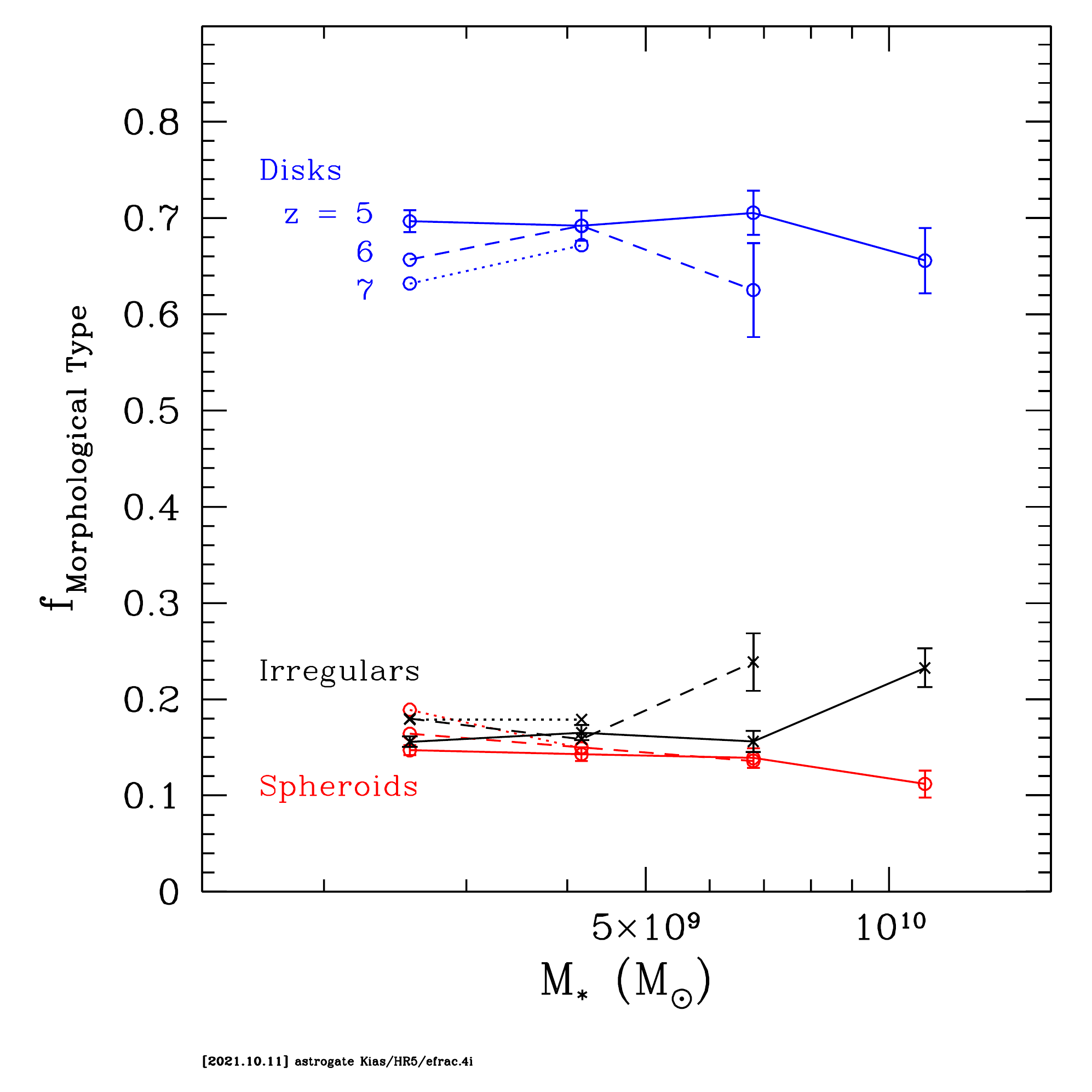}
\caption{Fractions of disk (blue), spheroid (red), and irregular (black) morphological types, at redshifts 5 (solid), 6 (dashed) and 7 (dotted) as a function of galaxy stellar mass.}
\label{fig:morphology_fraction}
\end{figure}

As described in Section 3, we have classified galaxy morphology based on the asymmetry parameter and the S\'{e}rsic index measured from the radial stellar-mass density profile. We measure the fraction of disks ($n_{\rm S\acute{e}rsic} < 1.5$), spheroids ($n_{\rm S\acute{e}rsic} > 1.5$), and irregulars ($A > 0.4$ or undefined $n_{\rm S\acute{e}rsic}$) in each mass bin at three redshifts. The dotted, dashed, and solid lines in Figure~\ref{fig:morphology_fraction} are the fractions at $z=7$, 6, and 5, respectively. It is manifest that the dominant morphological type of the galaxies in the cosmic morning is the disk type. The fraction of the disk type is between 0.63 and 0.69 for the galaxies with $M_{\star} > 2 \times 10^9 \,\msun$ at the three epochs. The fraction of the spheroid type is also quite stable and is between 0.19 and 0.14 for the same mass range. Interestingly, the fraction of irregulars is not high and is only a little higher than that of spheroids. However, the irregular fraction increases at the highest-mass end at each redshift, which indicates that the relatively more massive galaxies are likely to be undergoing structure-destroying interactions.

From figure~\ref{fig:morphology_fraction}, we conclude that the fraction of disk plus irregular morphological types does not depend much on stellar mass over the range we have explored. For example, the fraction is $\sim$0.87 within the range of $M_{\star} = 2-30 \times 10^9 \,\msun$ at $z=5$. On the other hand, the observed fraction decreases sharply for local galaxies from $\sim0.85$ to $\sim0.25$ over the same mass range~\citep[Figure 4 of][]{kelvin14}. This fact may imply that at high redshifts the mechanism determining galaxy morphology is different from the local one. 
The fraction of spheroid-type galaxies ($\sim0.14$ at $z=5$) found in this study is in good agreement with the fraction of early-type galaxies observed at $z \approx 1$ as measured by \citet{hwang09}, who have found using the GOODS data that the fraction converges to $\sim0.15$ in both high and low density regions near $z=1$~\citep[Figure~\ref{fig:morphology_fraction} in][]{hwang09}. %They  reported that the morphology-density relation has been developed after $z=1$ and starts to disappear at $z \gtrsim 1$, and that the galaxy-galaxy and galaxy-cluster core interactions below $z=1$ seem to be responsible for the relation. 
We note that the observed early-type fraction of about 0.15 at $z \approx 1$ has  already been set up in \hr\ at high redshifts even though the early types at $z>4$ are most probably not those at $z\approx 1$ according to our finding of frequent morphology transformation of high-redshift galaxies presented in the next section.
%(We will show in the forthcoming paper that the spheroid fraction stays at about 0.14 all %the way down to $z<2$ for galaxies with $M_{\star} < 5 \times 10^{10} \,\msun$ in \hr5.)
%[Most massive galaxies at a given epoch are undergoing structure-destructing interactions %more frequently relative to less massive ones.]******

Using the EAGLE simulation \citet{trayford19} has reported the distribution of high-redshift galaxy morphology very different from ours. For example, Figure 6 of the paper shows that irregulars (i.e. asymmetric galaxies) dominate at high redshifts and all the way down to $z\approx 1.4$. The difference comes from their low choice of asymmetry parameter cut of $A=0.2$ for irregulars. The choice is made from a visual inspection of galaxy images in their Figure 3. In the current study the distribution of measured $A$ is used to make the choice and the chosen threshold separates the disturbed galaxies from the large group of normal galaxies as seen in Figures 3 and 4. To some extent the division of irregular galaxies is a matter of subjective choice. In Trayford et al. the fractions of disks, spheroids, and irregulars are about 14, 24, and 63\% at z=6, which are very different from our results. It should be pointed out that their morphology classification of disk and spheroid types is actually based not on galaxy stellar mass distribution but on kinematics parameterized by the circularity $\epsilon$. As we show throughout this paper, the relation between morphology and internal kinematics is far from tight in the cosmic morning.

\subsection{Effects of Evolution}

\subsubsection{Evolution of S\'{e}rsic index of individual galaxies}

\begin{figure}
\centering 
\includegraphics[width=0.47\textwidth]{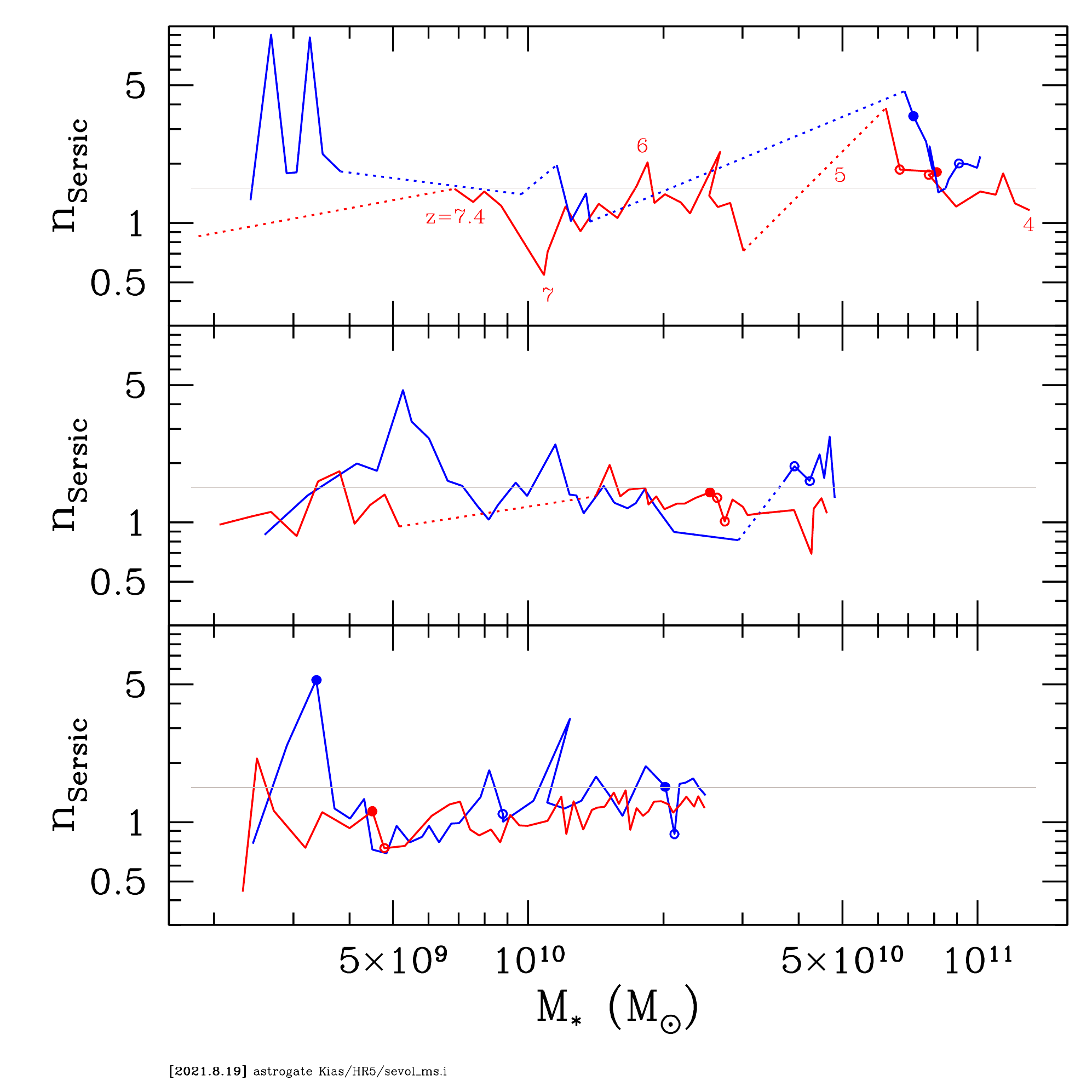}
\caption{Evolution tracks of galaxies in S\'{e}rsic index versus stellar mass plane. The top panel shows the evolution of two most massive galaxies at $z=4$. The middle and bottom panels exhibit the tracks of the galaxies with $M_{\star} \approx 5$ or $2.5 \times 10^{10} \,\msun$ randomly chosen at $z=4$. All the galaxies are traced after they reach $M_{\star}=2 \times 10^9 \,\msun$. The dotted lines mark the periods when galaxy morphology is irregular. Also shown are the time steps when the mass of the most massive super massive blackhole among the black holes belonging to a galaxy increases more than 50\% (filled circles) or 25\% (open circles) per $2.5\times 10^7$ years.}
\label{fig:evolution_track}
\end{figure}

By closely examining the galaxies in \hr, we find that the spheroidal and irregular morphologies of  galaxies in the cosmic morning are incidental and transient phenomena triggered by galaxy-galaxy interactions. To illustrate this, we show in Figure~\ref{fig:evolution_track} the evolution of the S\'{e}rsic index of a few galaxies as a function of their stellar mass through $z=4$. The red and blue lines in the top panel are the tracks of two most massive galaxies identified at $z=4$ after their stellar mass reached $2 \times 10^9 \,\msun$. The dotted lines indicate the periods when the S\'{e}rsic index is not measurable or the asymmetry parameter is greater than 0.4. For the most massive galaxy (red line) some specific redshifts are marked. The middle and bottom panels show the tracks of two randomly selected galaxies with $M_{\star} \approx 5$ and $2.5 \times 10^{10} \,\msun$ at $z=4$, respectively. The grey horizontal lines ($n_{\rm S\acute{e}rsic}=1.5$) are the demarcation dividing spheroids and disks in our study.

It can be immediately seen that galaxy morphology in the cosmic morning is not a fixed property but fluctuates in time. Galaxies spend most of their lives as disk types, but frequently change their morphology to spheroid or irregular type. One interesting phenomenon to note is that galaxies tend to recover their disk morphology after they become spheroids or irregulars. This tendency is likely to be due to re-acquisition of angular momentum from the vorticity-rich surrounding material~\citep[see, e.g.][]{Laigle2015,Danovich2015}, as we discuss below.

The S\'{e}rsic index of relatively low mass galaxies with $M_{\star} \leq 5 \times 10^{10} \,\msun$ at $z=4$ stays under the division line most of the time, and only occasionally upcrosses the line. The tracks of the more massive galaxies often have dotted lines indicating the periods of significantly perturbed or destructed structure. Some of them are associated with merger or breakup events resulting in abrupt mass changes, but not always so. When galaxies undergo destructive interactions resulting in irregular morphology, S\'{e}rsic index tends to increase. Among the galaxies with $M_{\star} = 5 - 25 \times 10^{9} \,\msun$ at $z=6$ that have suffered from structure-destructing events $56 \pm 4\%$ have increased their S\'{e}rsic indices after each event while $44 \pm 3.7\%$ have decreased it. On the other hand, it is nearly equally likely for S\'{e}rsic index to increase or decrease when all major stellar mass jumps, $\Delta M_{\star} /M_{\star} > 0.25$ per simulation output time step, are considered regardless of morphology transformation to irregular type.

\begin{figure}
\centering 
\includegraphics[width=0.45\textwidth]{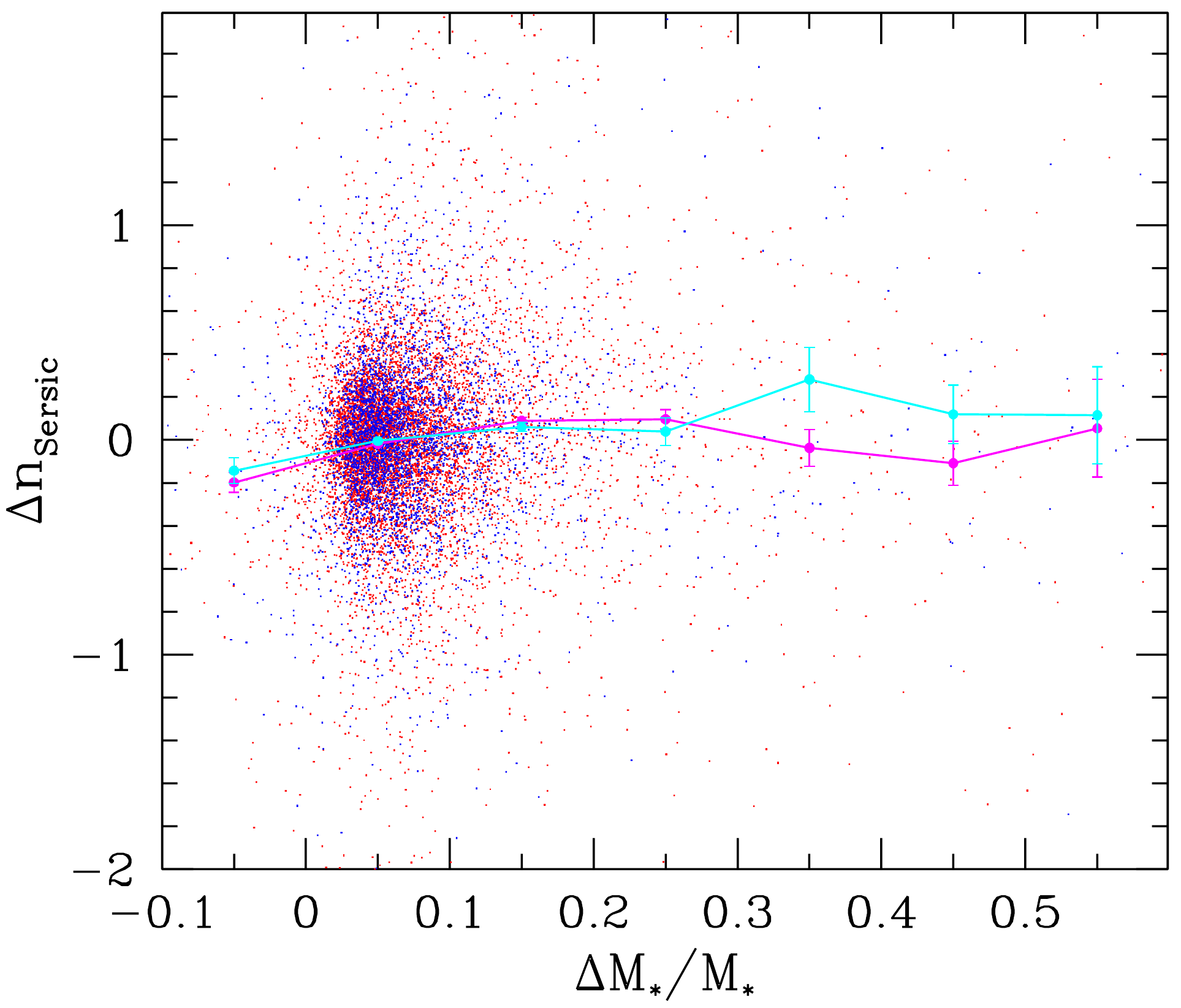}
\caption{S\'{e}rsic index change  $\Delta n_{\rm S\acute{e}rsic}$ versus  fractional stellar mass change $\Delta M_{\star}/M_{\rm star}$ per unit output time step of $\Delta t\approx 2.5\times10^7\,$yr. Each dot corresponds to a particular time step between $z=4$ and the epoch when each galaxy first reaches $M_\star=2\times10^9\,\msun$. The blue dots indicates the galaxies with $M_{\star} = 5 - 6 \times 10^{9} \,\msun$ and the red dots are for those with $M_{\star} > 6 \times 10^{9} \,\msun$ at $z=6$. The cyan and magenta lines show the mean values of $\Delta n_{\rm S\acute{e}rsic}$ for blue and red points, respectively. The error bars exhibit $1\sigma$ distribution of the mean curves.}
\label{fig:sersic_change}
\end{figure}

Figure~\ref{fig:sersic_change} shows the change in S\'{e}rsic index of galaxies and fractional stellar mass change that occur at each output time step of $\Delta t\approx2.5\times10^7$ yr. The time steps between $z=4$ and the epoch when a galaxy first reaches $M_{\star}=2\times 10^9\,\msun$  are used for the plot. The upper redshift limit thus varies among galaxies. The galaxies that have $M_{\star} = 5-6 \times 10^{9} \,\msun$ (blue points) or $M_{\star} > 6 \times 10^{9} \,\msun$ (red points) at $z=6$ are selected and traced along the merger tree. Cyan and magenta lines are the mean values for the two mass intervals, respectively, and error bars are the uncertainties of the means. S\'{e}rsic index shows no trend with mass change when $\Delta M_{\star} /M_{\star} > 0.15$. We conclude that the destructive interactions resulting in spheroids or irregulars are not always identical to merger events. On the other hand, there is a very weak but statistically significant positive relation between $\Delta M_{\star} /M_{\star} =0.05$ and $0.15$ (more than 6 $\sigma$). This means that, when $\Delta M_{\star} /M_{\star} \lesssim 0.15$, galaxies on average tend to make more positive S\'{e}rsic index change for larger fractional stellar mass increase even though the change at each step appears stochastic. This is the regime of internal stellar mass growth due to star formation within galaxies or accretion of small mass objects. The star formation rate of the galaxies with $M_{\star} =5 \times 10^{9} \,\msun$ at $z=6$ is typically about $20 \,\msun$/year and they increases stellar mass of about $5\times 10^8 \,\msun$ or $\Delta M_{\star} /M_{\star} \approx 0.1$ during $2.5\times 10^7$ years. 

In Figure~\ref{fig:evolution_track}, we mark the time steps when the mass of the most massive super massive blackhole (SMBH) among the black holes belonging to a galaxy increases more than 50\% (filled circles) or 25\% (open circles) per $2.5\times 10^7$ years. Studies with much larger samples need to be made to draw any conclusions whether or not sudden growth of SMBH mass is correlated with morphological transformation at high redshifts.

\begin{figure}
\centering 
\includegraphics[width=0.45\textwidth]{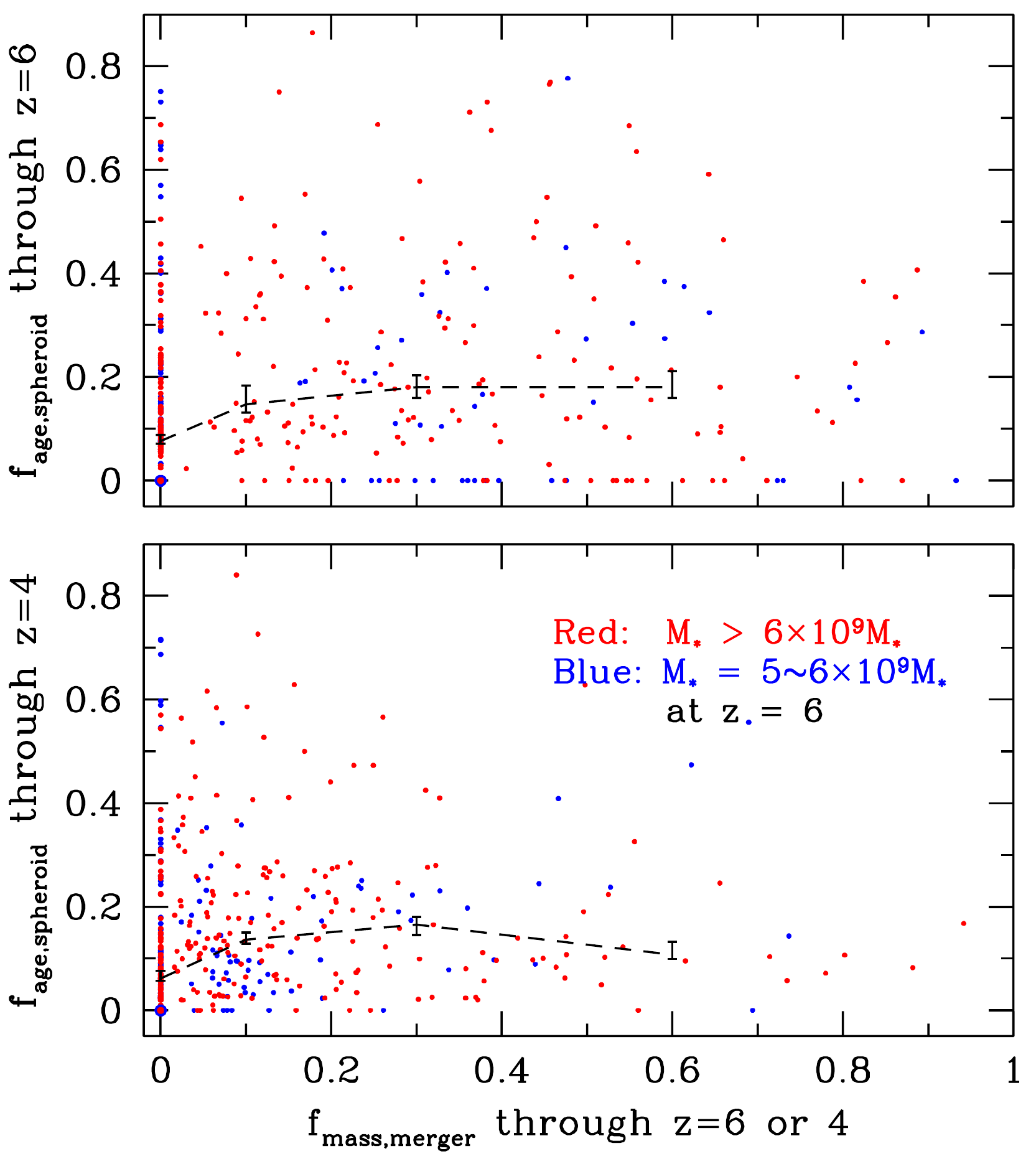}
\caption{Fraction of age that galaxies have spent their lives as spheroids by $z=6$ (top) and 4 (bottom) after their stellar mass reaches $M_{\star}=2\times10^9\,\msun$. The abscissa is the fraction of stellar mass {red increased through mass jumps typically accompanied by mergers (see the text)}. The blue and red dots indicate galaxies having $M_{\star}=5-6\times10^9\,\msun$ and $M_{\star} >6\times10^9\,\msun$ at $z=6$, respectively. There are 44 (20) blue and 38 (18) red points at the origin out of a total 278 (146) points in the upper (lower) panel. The dashed curves present the medians at given $f_{\rm mass,merger}$, and error bars are the uncertainties of the medians.}
\label{fig:fage_morphology}
\end{figure}

\subsubsection{Effective morphology}

Since galaxy morphology at a particular epoch cannot represent the shape of galaxies over an extended period, we instead introduce the fraction of the time period during which a galaxy has spent its life as a spheroid, to assign a representative galaxy morphology. That is, we define
\begin{equation}
 f_{{\rm age,spheroid}}(z) \equiv \frac {\displaystyle\int_{t_i}^{t(z)} \Theta (n_{\rm S\acute{e}rsic}(t) - 1.5) dt } {t(z)-t_i},
\end{equation}
where $t_i$ is the time when the galaxy reached stellar mass of $2\times 10^9 \,\msun$, and $\Theta$ is the step function. Similarly, we define the fraction of stellar mass that has increased through mergers by a given redshift $z$,

\begin{equation}
f_{{\rm mass,merger}}(z) \equiv  \frac{\displaystyle \int_{z_i}^{z} \Theta (\Delta M_{\star} /M_{\star} - 0.25) dM_{\star}(z') }{M_{\star}(z)-M_{\star}(z_i)},
\end{equation}
where the mass change $\Delta M_{\star}$ is again measured for our output time steps. 
Here the stellar mass increase $\Delta M_{\star}$ not only includes the stellar mass directly carried to the major galaxy by the infalling minor one but also that immediately formed after merger events. Therefore, the criterion $\Delta M_{\star} /M_{\star} > 0.25$ per time step does not always find the conventional major mergers used to construct the merger tree but also includes some minor mergers that trigger strong star burst.

Figure~\ref{fig:fage_morphology} shows the fraction of time that galaxies live as spheroids from the time when they acquired stellar mass of $2\times 10^9 \,\msun$ through $z=6$ (upper panel) or 4 (bottom panel). The abscissa is $f_{\rm mass,merger}$ integrated through the corresponding redshifts. Blue and red points are again the galaxies having stellar mass of $5 - 6\times10^9\,\msun$ or greater than $6 \times 10^{9} \,\msun$ at $z=6$, respectively. They indicate that both effective morphology and merger-driven stellar mass fraction do not depend on the stellar mass of galaxies. Note that there are many points at $(0,0)$ overlapping one another. They are the galaxies that have never had spheroidal morphology and never experienced major stellar mass jumps. Dashed curves are the median $f_{\rm age,spheroid}$. 
%as a function of $f_{\rm mass,merger}$.

$f_{\rm mass,merger}$ is on average a little higher in the upper panel, because at the beginning of cosmic morning galaxies more frequently merge with other galaxies with similar mass. As can be seen from the median curves, the majority of galaxies spend their lives as late types (i.e. disks or irregulars) and the probability for them to live as spheroid is essentially independent of $f_{\rm mass,merger}$ as long as $f_{\rm mass,merger}$ is not zero. Interestingly, the galaxies that have never experienced major mass jumps %tend to have non-spheroid morphology. %For example, these galaxies 
have $f_{\rm age,spheroid} = 0.077^{+0.011}_{-0.005}$ at $z=6$, which is somewhat lower than $f_{\rm age,spheroid} = 0.169^{+0.030}_{-0.019}$ for those that have ever experienced major mass jumps.  

In most cases our statistical study includes the galaxies with $M_{\star} = 2 -10 \times 10^9 \,\msun$ in the cosmic morning ($z > 4$). Our results suggest that these galaxies are typically disks and temporally become irregulars or spheroids. The mechanism transforming galaxies from disks to spheroids on average is not simply mass-jumping merger events, as such mergers do not preferentially make galaxies more spheroidal in this mass range. Instead, a statistical trend for transformation to spheroidal type is observed when galaxies go through an irregular morphology phase. Various kinds of cataclysmic events like close encounters, major and minor mergers, and episodic strong accretion that manage to affect galaxy structure significantly seem responsible for the trend. \cite{welker14} has showed that the magnitude of the spin of simulated galaxies increases steadily and aligns itself preferentially with the nearest filament when no significant merger occurs. Strong internal star formation gives a weak tendency toward spheroidal type too~\citep{Martin2018}. 

%\subsubsection{Star Formation and Structure-Destructing Events}

\subsection{Effects of the Initial Conditions}

\begin{figure}
\centering 
\includegraphics[width=0.47\textwidth]{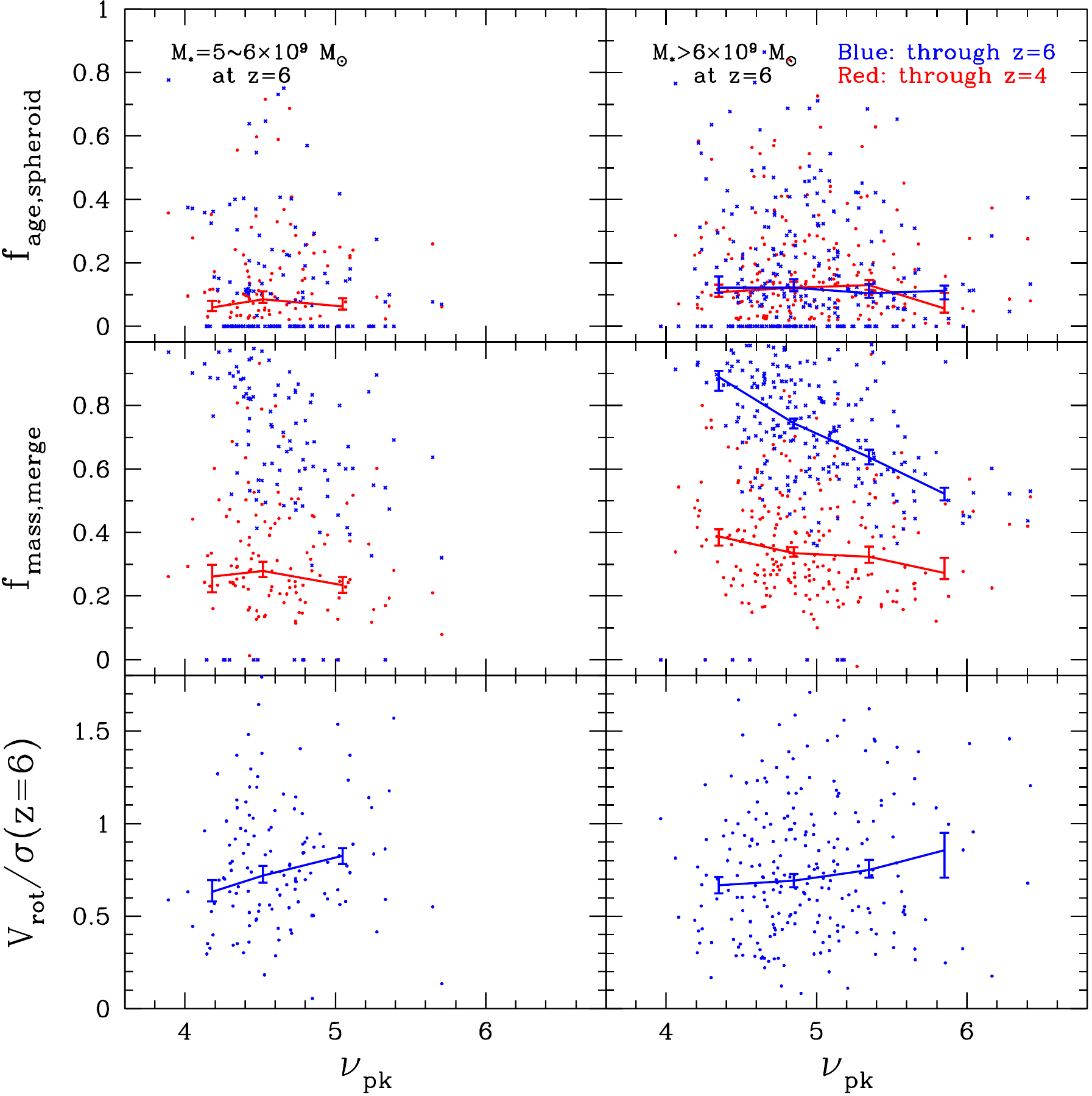}
\caption{Fraction of age that galaxies spend their lives as spheroids (top), fraction of stellar mass increased during mergers (middle), and rotational velocity to velocity dispersion ratio at $z=6$ (bottom) as a function of the height $\nu_{\rm pk}$ of the initial dark matter overdensity peak matched to galaxies. Density peaks are defined on the `galaxy' scale of $R_G=0.35$ cMpc. The blue and red dots in the top and middle panels are the fractions integrated through $z=6$ and 4, respectively, after their stellar mass reaches $2\times10^9\,\msun$. The left panels are for the galaxies with  $M_{\star}=5-6\times10^9\,\msun$ and the right panels for those with $M_{\star} > 6\times10^9\,\msun$ at $z=6$. Median curves and their uncertainty limits are given.}
\label{fig:fage_fmass_rot_density}
\end{figure}

It is expected that the monolithic collapse of mass around initially very high density region ends up producing galaxies morphologically different from those formed in filaments or walls. It is also sensible to expect for kinematically different regions that are in different tidal force environment~\citep[][their Fig~12]{codis15} to produce morphologically different galaxies. In this section we examine how  the initial conditions impact galaxy properties in the cosmic morning.  

\subsubsection{Effects of the initial peak height}

We inspect in Figure~\ref{fig:fage_fmass_rot_density} if the initial density peak height affects galaxy mass, morphology, merger history, and kinematic property of galaxies. The abscissa is the height of the initial density peak matched with the initial center of mass of the dark-matter particles composing the dark matter halo of each galaxy identified at $z=6$.
The initial dark matter density field represented on a 0.128 cMpc pixel mesh and Gaussian smoothed over $R_G =0.35$ cMpc is used to identify the initial density peaks and to measure their height $\nu_{\rm pk}\equiv(\rho_{\rm DM}/\bar{\rho}_{\rm DM}-1)/\sigma_{\rm DM}$, where $\sigma_{DM}$ is the RMS density fluctuation. 
Plotted points are for the galaxies that have $M_{\star} = 5 - 6 \times 10^9 \,\msun$ (left panels) or $M_{\star} > 6 \times 10^9 \,\msun$ (right panels) at $z=6$. The blue and red points in the top and middle panels are the fractions calculated through $z=6$ and 4, and the red/blue lines with error bars are the medians of the red/blue dots, respectively. 

In the middle panels, $f_{{\rm mass,merge}}$ is the fraction of galaxy stellar mass that has increased during mass jumps. Here we lower the threshold for the mass jump to $\Delta M_{\star} /M_{\star} =0.1$ per time step to include more minor mergers, so it is defined somewhat differently from that in Figure~\ref{fig:fage_morphology}. The threshold is lowered because the majority of galaxies have zero amount of mass acquired through major mass jumps, and it is difficult to see any trend with $\nu_{\rm pk}$. The stellar mass increase of 10\% during 0.025 Gyrs is the mass that galaxies at this epoch typically increase just through internal star formation. 

The top panels of Figure~\ref{fig:fage_fmass_rot_density} show that galaxy morphology is independent of the height of the initial density peaks that end up forming the first galaxies. This contrasts with the trends that  merger fraction and $V_{\rm rot}/\sigma$ do have some dependence on $\nu_{\rm pk}$ (middle and bottom panels). Higher initial density peaks tend to suffer from less amount of mergers by $z=6$ or 4. And higher density peaks tend to have higher $V_{\rm rot}/\sigma$ with a marginal statistical significance of $2.6 \sigma$. However, these trends in merger history and kinematic property are not related to galaxy morphology. We also do not see any clear dependence of morphology on age (blue versus red points) or mass (left versus right panel). Note that no galaxy in our sample has stellar mass higher than $3\times 10^{10} \,\msun$ before $z=6$ as can be seen in Figure~\ref{fig:prop_dm_density}.

\begin{figure}
\centering 
\includegraphics[width=0.42\textwidth]{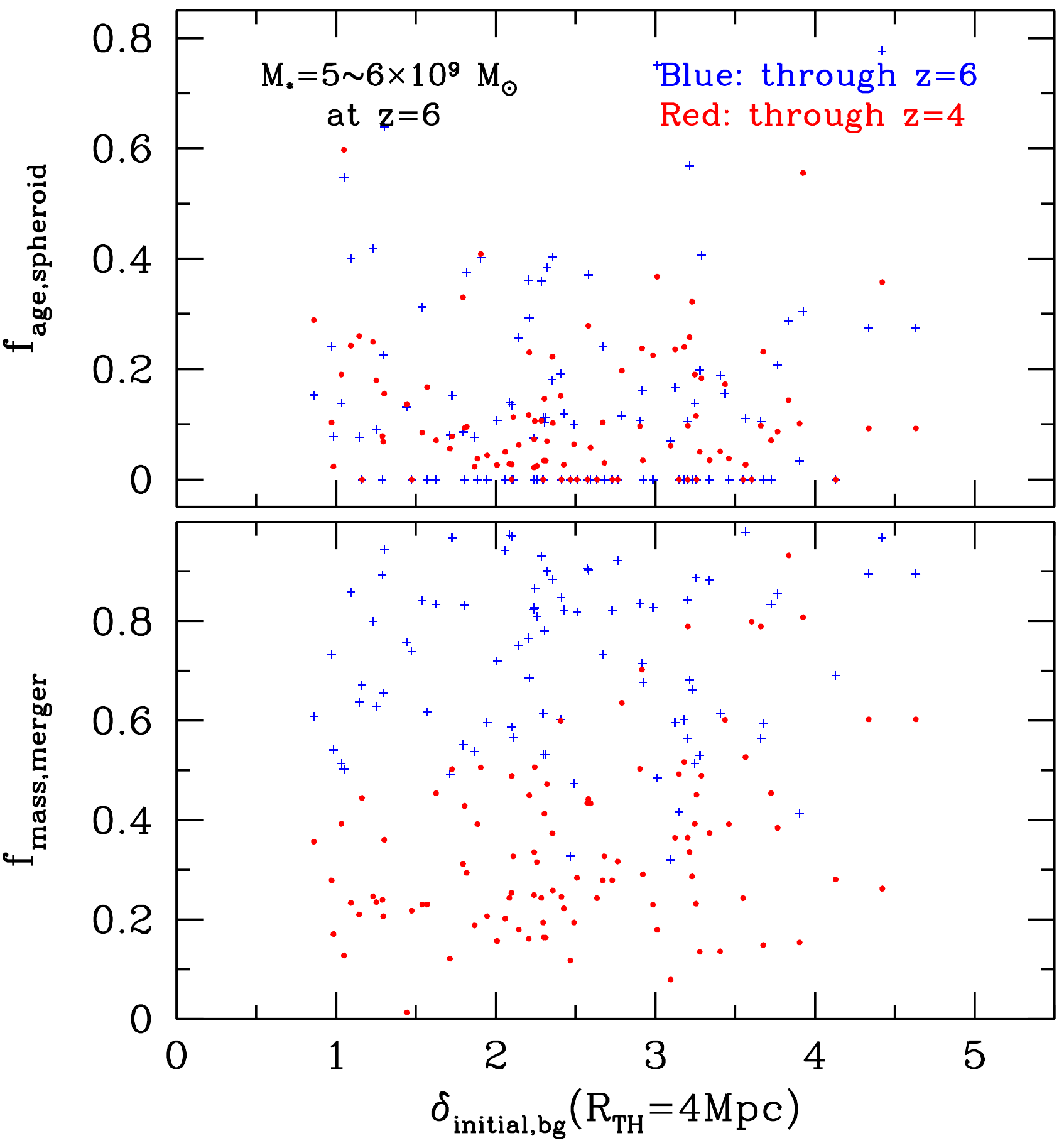}
\caption{Fraction of age that galaxies spend their lives as spheroids (top) and stellar mass fraction increased during mergers (bottom) integrated through $z=6$ (blue) and 4 (red) since galaxy mass reaches $M_{\star}=2\times10^9\,\msun$. The abscissa is the initial background overdensity contrast ($\delta_{\rm initial,bg}\equiv (\rho_{\rm DM}/\bar{\rho}_{\rm DM}-1)/\sigma_{DM}$) smoothed with a top-hat filter over the `proto-cluster' scale of $R_{\rm TH}=4$\,cMpc.}
\label{fig:fage_fmass}
\end{figure}

\subsubsection{Effects of the large-scale background density}

To look into the dependency of galaxy morphology on large-scale density environment, in Figure~\ref{fig:fage_fmass} we plot $f_{\rm age,spheroid}$ and $f_{\rm mass,merger}$ as a function of the initial overdensity contrast $ (\rho_{\rm DM}/\bar{\rho}_{\rm DM}-1)/\sigma_{DM}$ smoothed with a top-hat filter over the `galaxy proto-cluster' scale of 4 cMpc. The corresponding smoothing mass is $1.0\times 10^{13}\,\msun$. We chose a sample of galaxies identified at $z=6$ having stellar mass within a narrow range of $5-6\times 10^9 \,\msun$. The fractions are integrated from the epoch when their stellar mass reached $2\times 10^9 \,\msun$ to $z=6$ (blue dots) or 4 (red dots). 

It can be seen in the upper panel that galaxy morphology is basically independent of the initial large-scale background density, and that spheroidal shape is a minority everywhere and late-type (disk plus irregular) morphology dominates both in high and low density regions in these early periods. The stellar mass fraction of galaxies acquired during mass jumps is also nearly independent of the initial background density by $z=6$. Galaxies grow mainly via mergers by that epoch as presented by the blue crosses in the lower panel. However, merger events decrease in time at different rates in different density environments, and by $z=4$ (stellar mass) accretion and internal star formation rather than mergers become the main channel of stellar mass growth of the galaxies in low background density regions (red dots). This environmental dependence of the merger/interaction rate is expected to be responsible for the environmental galaxy mass and morphology dependence at later epochs.

\subsubsection{Impact of the Initial tides}

Next we examine if galaxy morphology, merger fraction, and kinematics depend on the primordial angular momentum of galaxies. According to  tidal torque theory~\citep{doroshkevich70,white84}, the large-scale matter distribution can exercise tidal torque on a proto-galactic region, producing the angular momentum (hereafter AM) of the galaxy forming there. In the Zel'dovich approximation of Lagrangian perturbation theory~\citep{zeldovich70}, the AM of a mass distribution is given as follows~\citep{Sugerman2000,schafer12}:
\begin{equation}
\mathcal{L}_{\alpha}(\boldsymbol{q}_c) = a^3 H(a) \frac{dD}{da} \epsilon_{\alpha \beta \gamma} \sum_{\sigma} I_{\beta \sigma} \Psi_{\sigma \gamma}(\boldsymbol{q}_c),
\end{equation}
where $a$ is the scale factor, $H$ is the Hubble expansion rate, $D$ is the linear growth function that encodes the time evolution of the growing mode solution of linear density perturbation, $\epsilon_{\alpha \beta \gamma}$ is the Levi-Civita symbol, and $\boldsymbol{q}_c$ is the center of mass. The moment of inertia tensor in terms of the initial matter position $\boldsymbol{q}$ is
\begin{equation}
I_{\beta \sigma} = \rho_0 a^3 \int_V d^3 q (\boldsymbol{q}-\boldsymbol{q}_c)_{\beta} (\boldsymbol{q}-\boldsymbol{q}_c)_{\sigma},
\label{eq:momentum_inertia}
\end{equation}
where the integration is made over the proto-galactic volume $V$. The initial shear tensor is given by
\begin{equation}
\Psi_{\sigma \gamma}(\boldsymbol{q}_c) = \partial_{\sigma} \partial_{\gamma} \Psi(\boldsymbol{q}_c),
\label{eq:shear_tensor}
\end{equation}
where $\Psi(\boldsymbol{q}_c)$ is the displacement potential in the Zel'dovich approximation and we have assumed that the displacement field is irrotational. As we are interested only in the relative difference of AM, all the position-independent terms will be omitted in our evaluation of angular momentum. As the scalar displacement field is proportional to the gravitational potential in the linear approximation, we adopt a quantity proportional to the primordial AM as
\begin{equation}
L_{\alpha} \equiv \epsilon_{\alpha \beta \gamma} \sum_{\sigma} I_{\beta \sigma} \partial_{\sigma}\partial_{\gamma}\Phi (\boldsymbol{q}_c),
\label{eq:angular}
\end{equation}
where $\Phi$ is the gravitational potential field. We calculate the AM in the following way. We identify galaxies at $z=6$ and make a list of all dark-matter particles belonging to the dark-matter halo of each galaxy. We then use their initial positions to calculate the center of mass $\boldsymbol{q}_c$ and the moment of inertia tensor $I_{\beta\sigma}$. The gravitational potential field is smoothed with a Gaussian filter over $R_G = 0.35$ cMpc, and the {\it shear} tensor $\partial_\sigma\partial_\gamma\Phi$ is calculated at $\boldsymbol{q}_c$ by using the finite difference method. The moment of inertia tensor and the gravitational shear tensor are used to calculate the AM in Equation~\ref{eq:angular}.

\begin{figure}
\centering 
\includegraphics[width=0.475\textwidth]{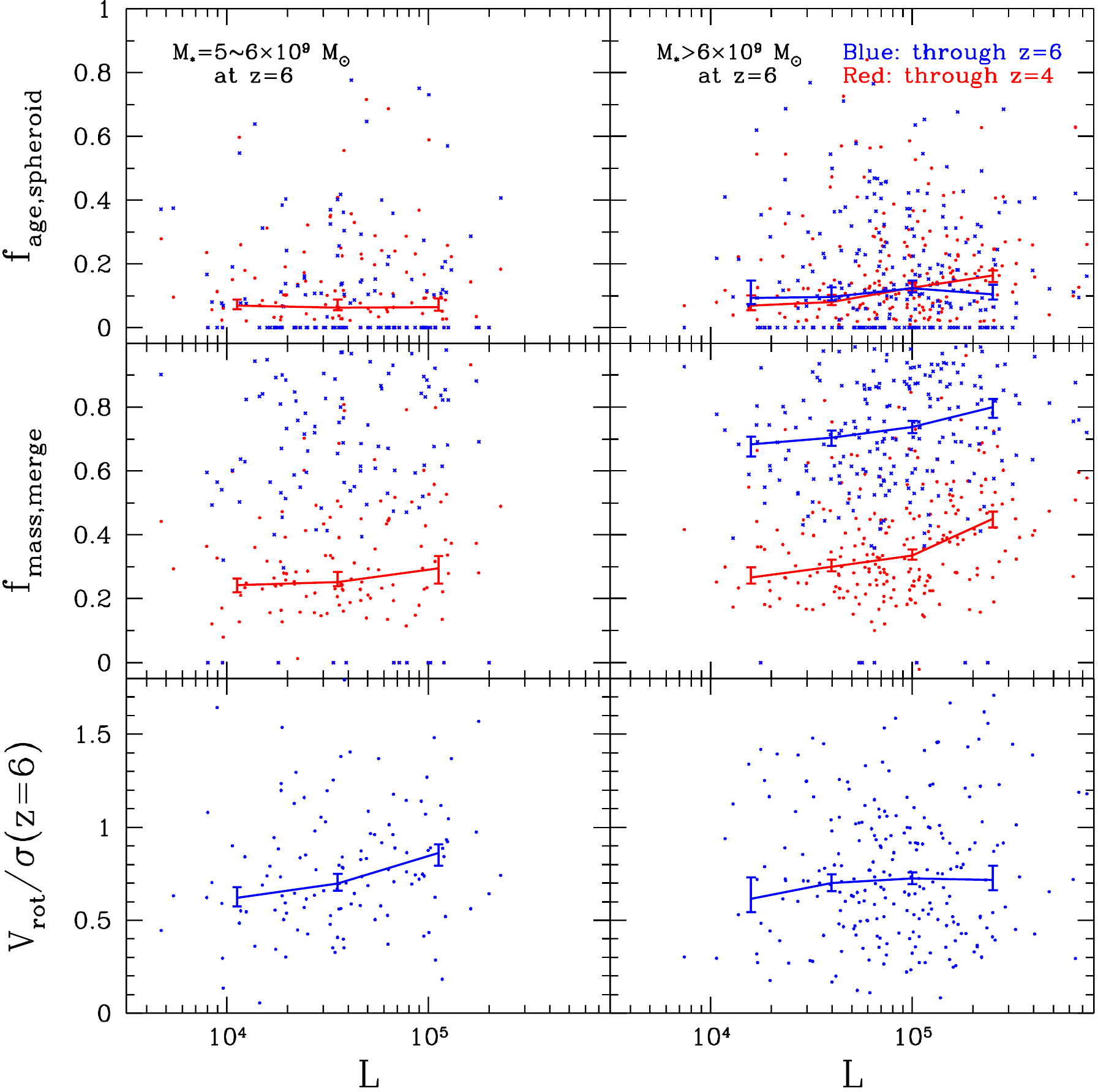}
\caption{Fraction of age that galaxies spend their lives as spheroids (top), stellar mass fraction increased during mergers (middle), and rotational velocity to velocity dispersion ratio of stars at $z=6$ (bottom) as a function of the initial angular momentum (in arbitrary unit). Blue and red points in the top and middle panels are the fractions integrated through $z=6$ and 4, respectively, after their stellar mass reaches $2\times10^9\,\msun$. Left panels are for the galaxies with  $M_{\star}=5-6\times10^9\,\msun$, and the right panels for those with $M_{\star} > 6\times10^9\,\msun$ at $z=6$.}
\label{fig:fage_fmass_rot_ang}
\end{figure}

\begin{figure}
\centering 
\includegraphics[width=0.47\textwidth]{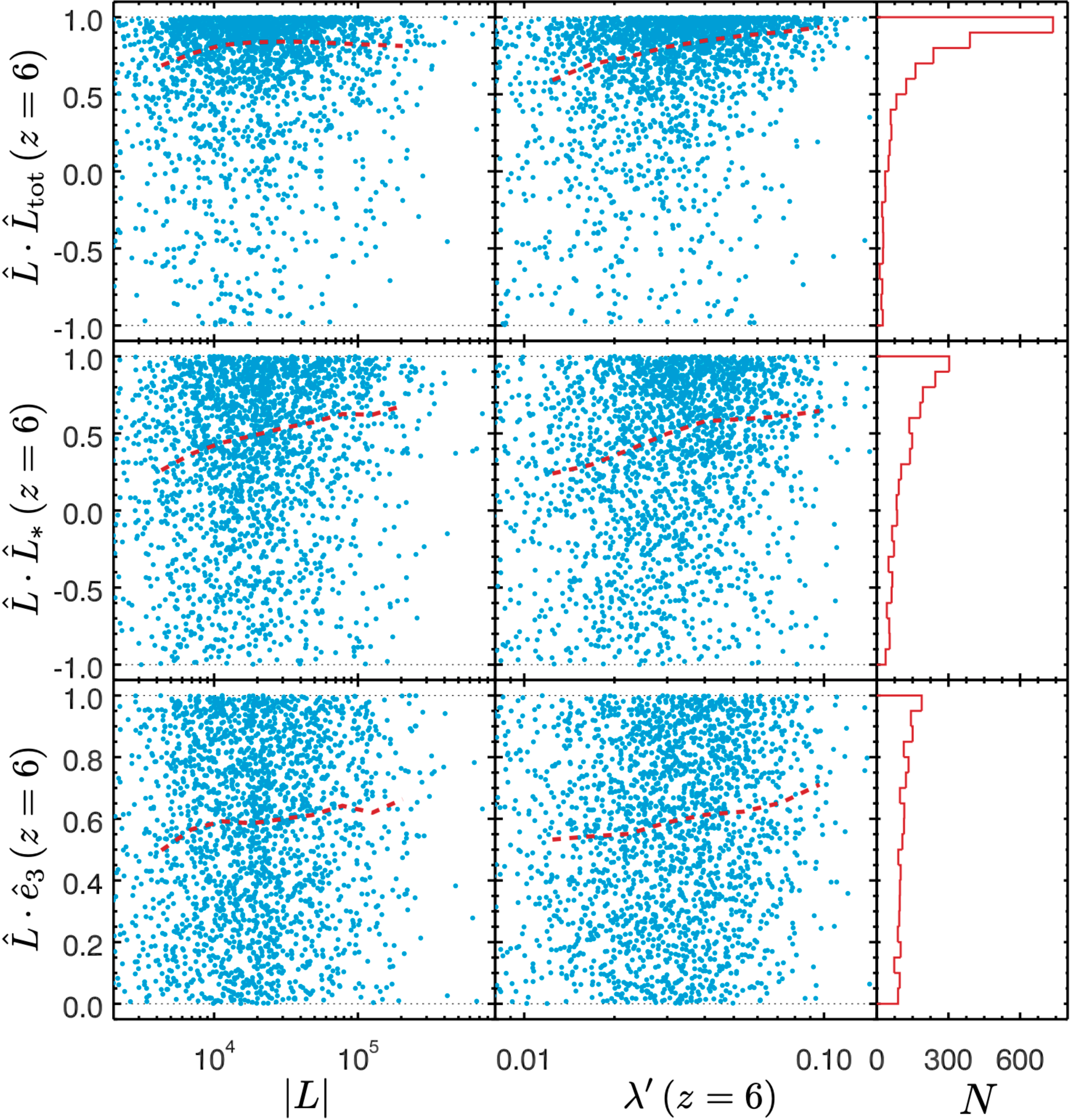}
\caption{Alignment between initial angular momentum and angular momentum of total mass (top panels), angular momentum of stellar particles (middle) and the minor axis of  the moment of inertia of the stellar mass distribution (bottom) as functions of the magnitude of initial angular momentum (left column) and spin parameter $\lambda'$ (middle column). Histograms are shown on the right. Galaxies with  $M_{\star} > 2\times10^9\,\msun$ at $z=6$ are used. Initial angular momentum is strongly aligned with the total angular momentum of galaxies. }
\label{fig:spin_align}
\end{figure}

The top panels of Figure~\ref{fig:fage_fmass_rot_ang} show that galaxy morphology is basically independent of the initial AM of proto-galactic regions. That is,  disk morphology dominates regardless of the magnitude of $L$. Here $f_{\rm age,spheroid}$ is nearly independent of $L$ both for relatively low and high mass galaxies, even though the fraction seems to develop a moderately increasing trend with $L$ between $z=6$ and 4 for massive galaxies (i.e. red line in the top right panel). 
Even though the galaxy morphology does not show dependence on the initial AM, $f_{\rm mass,merge}$ and $V_{\rm rot}/\sigma$ of stars are weakly increasing function of $L$ as shown in the middle and bottom panels. 
%It should be noted however that the dependence of internal kinematics on $L$ is weak. An order of magnitude variation in angular momentum makes $V_{\rm rot}/\sigma$ change only a little as shown in the bottom panels. This may indicate that  *********

%Comparison between the left and right middle panels leads us to conclude that initially higher angular momentum regions are the sites where more massive galaxy form and mass grows more dominantly through mergers. On the other hand, initially higher density peaks form higher mass galaxies that undergo less amount of mergers (top panels of Figure~\ref{fig:prop_dm_density} and middle panels of Figure~\ref{fig:fage_fmass_rot_density}). These apparently conflicting trends are related with the appearance of the massive galaxies with $M_{\star} > 3\times 10^{10} \,\msun$ after $z=6$, which will be discussed in Discussion section below. The weaker dependence on $V_{\rm rot}/\sigma$ on $L$ for more massive galaxies (bottom right panel) seems to be also related with mergers of massive galaxies that are losing their initial memory.

Evidence for a positive correlation between the initial AM and internal kinematics and  for the negligible correlation between $L$ and galaxy morphology can also be  seen in Figure~\ref{fig:spin_align}. The right-most panels of Figure~\ref{fig:spin_align} presents histograms showing the alignment of the initial AM with the AM of total mass ($L_\mathrm{tot}$, top panel), stellar component ($L_{\star}$, middle panel) and with the minor axis of the moment of inertia of stellar mass distribution (${\hat e}_3$, bottom panel) of galaxies at $z=6$.
It can be seen that the primordial AM is tightly aligned with the total AM, and rather well aligned with the AM of the stellar component. ${\hat L} \cdot {\hat L}_{\rm galaxy}({\rm z=6})$ is greater than $0$ for 88.3\% or 73.6\% of galaxies when total (dominantly dark matter) or stellar mass is used for AM calculation, respectively. However, ${\hat L}$ shows only very weak alignment with the short axis of the stellar mass distribution. It should be pointed out again that the initial $L$ is tightly aligned with galaxy internal motion but not much aligned with the mass shape. This suggests that one should not use internal kinematics to define galaxy morphology at high redshifts.

The spin parameter $\lambda'$  of galaxies at $z=6$ in Figure~\ref{fig:spin_align} is defined as
\begin{equation}
\lambda' = {\mathcal{L} \over {\sqrt{2} MvR}},
\label{eq:lambda}
\end{equation}
where $\mathcal{L}$ is the total galaxy AM, $M$ is virial mass, $R$ is virial radius, and $v$ is circular velocity~\citep{bullock01,Aubert2004,knebe08,schafer12}.
The alignment is somewhat stronger for galaxies with higher $L$ or $\lambda'$, as shown by the red dashed median lines. 

A possible interpretation is that the short axis of stellar mass basically shows nearly no alignment with the initial AM because the morphology of stellar component of galaxies in the cosmic morning can be easily and significantly affected by  perturbations, hence has a limited memory of the initial conditions.
On the other hand, the AM of stellar mass of galaxies does show some alignment, and the AM of the total mass has even stronger alignment with the initial AM. As the (gas dominated) mass reservoir of galaxies remembers the initial AM produced by the tidal torque of large-scale dark matter distribution, later in-falling gas onto the galaxies tends to reset the original disk morphology~\citep[e.g.][]{Pichon2011,Stewart2013}.

\section{Discussion}

\begin{figure*}
\centering 
\includegraphics[width=0.73\textwidth]{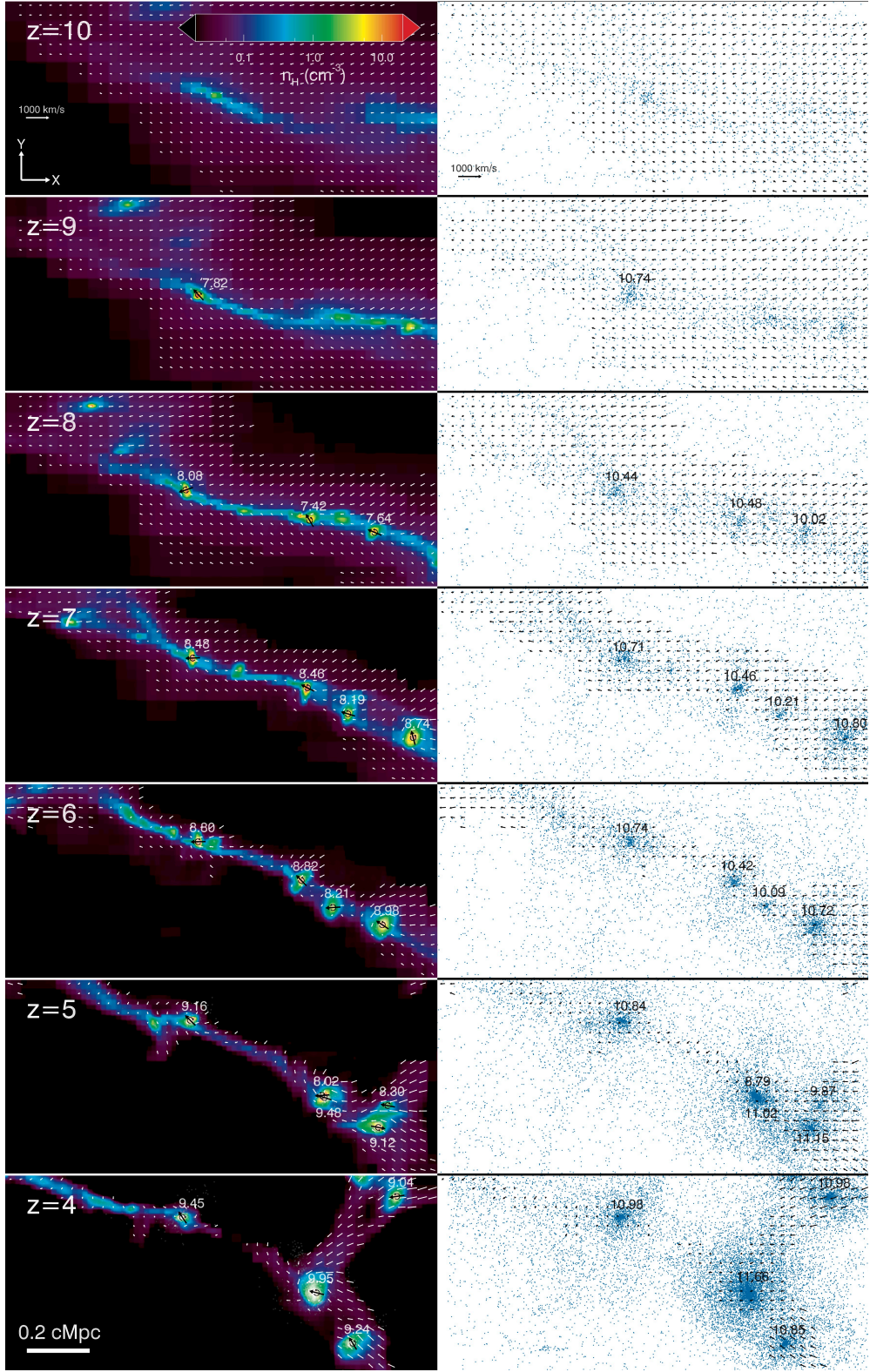}
\caption{An 1.5 cMpc-thick slice of a relatively high density region in HR5 showing gas density and stars (left) and dark matter (right) from $z=10$ to 4.
%on an X-Y plane. 
Gas density is higher in redder shades. Whitish grey dots in the left illustrate stellar mass distribution. The white arrows show the velocity field of gas in dense areas. The black arrows at dense cores mark the spin direction of galaxy stellar mass projected on the slice. The spin direction perpendicular to the slice (clockwise or counter-clockwise rotation) is also shown on top of the spin arrows.
%Galaxies rotate counter-clockwise on the projected plane when the arrow heads penetrate the open circles, i.e., the galaxies have positive spin components in the z-direction. On the other hand, the open circles penetrated by the tail of the arrows indicate clockwise rotation of galaxies on the plane. 
Numbers are logarithmic stellar mass (left) and dark matter halo mass (right) in $M_{\odot}$ unit. %The black arrows in the right panels display the velocity field of dark matter.
The areas where dark matter velocity field is shown are the same as in the gas density plot.}
\label{fig:filaments}
\end{figure*}

In this work, we have found that the majority ($\sim2/3$) of galaxies in the cosmic morning before $z=4$ have a disk morphology, and galaxies transiently take a spheroidal or irregular morphology when they experience destructive interactions. 

This finding, in broad agreement with previous studies, raises many questions: i)  why should the default initial morphology be a disk? ii) what makes the fraction of the transient morphology (i.e. spheroid or irregular type) a particular value of about 1/3? iii) why does the fraction not change much with time or stellar mass? and iv) why do merger events not always preferentially result in spheroidal morphology, as found for the massive galaxies at low redshifts? The inspection of the relations between the galaxy morphology and the initial conditions leads us to raise more questions such as v) when does the environmental dependence of galaxy morphology appear? and  vi) when is the correlation between morphology and kinematics (re-)established?

While we cannot answer all of these questions in view of this preliminary work on \hr, let us address specifically a few of them here. 
One key ingredient seems to be the consistency of gas-rich inflow, set up by the larger cosmic scales, via a top-down causation.
Galaxy morphology is disk dominated, but is poorly correlated with the initial angular momentum (AM) as seen in Figures~\ref{fig:fage_fmass_rot_ang} and \ref{fig:spin_align}. The initial AM produced by the tidal torque on protogalactic regions tends to make newly born galaxies disk-like~\citep{navarro04,Bailin2005,Pichon2011,codis12}, making disks the dominant morphology in the cosmic morning. 
Indeed, the dynamics within neighbouring rhizomes drives cold gas out of the surrounding voids, into accreting filaments. 
As boundaries of asymmetric voids, they acquire a net transverse motion, which explains the angular momentum rich nature of infall. This large-scale driven consistency explains why cold streams are so efficient at building up cosmic morning disks.
The tight correlations of the AM and spin parameter of the stellar component of galaxies with the initial AM found in this simulation support this interpretation. 

It is also expected that structure formation and evolution on a galactic scale have been largely controlled by the large-scale power of the tidal torque,
%within the surrounding cosmic web, 
which has played an important role in setting up the preferred direction of  inflow~\citep{Sugerman2000,Lee2000,Aubert2004,Bailin2005}. The total mass of galaxies having $M_{\star} =1-2\times 10^9\,\msun$ at $z=6$ is typically $M_{\rm tot}=1\times 10^{11}\,\msun$. The radius of tophat sphere containing this much of mass in the initial condition is $R_{\rm TH}=0.849$ cMpc in the cosmology adopted in this paper. The slope of our $\Lambda$CDM power spectrum is $d {\rm ln} P / d {\rm ln} k =-2.63$ at $k=2\pi/R_{\rm TH}$. %At such a negative slope scale 
When the slope of power spectrum is close to $d{\rm ln}P/d{\rm ln}k = -3$, the structure formation proceeds almost at the same time from sub-galactic to super-galactic scales, and influence of large-scale modes on galaxy formation should therefore be significant, leading to a top-down impact of the larger scale on galactic infall~\citep{zeldovich70b} ~\citep[, see also][for a complementary geometric formulation in terms of the size of vorticity and AM caustics  and the mass of non-linearity at that  redshift]{codis12,codis15,Laigle2015}. 
The situation is quite different for galaxy clusters where the slope of power spectrum at the relevant scale is about $-1$. The first non-linear `galactic-scale' structures appearing at  cosmic dawn ($z \gtrsim 10$) are isolated spots corresponding to the highest peaks in the initial density field, smoothed over a few hundred ckpc. Nearly simultaneously to the collapse of the peaks, the surrounding matter also collapses to form a few non-linear filaments attached to them. Rapid mass accretion occur onto  peaks and filaments~\citep[along the  cold flows,][]{Keres2005,Dekel2006,Ocvirk2008M}.  

Figure~\ref{fig:filaments} shows zoomed snapshots of a relatively high density region with 1.5 cMpc thickness in HR5 where some first galaxies and filaments are forming. Left column shows gas density and stars, and right column shows dark matter particles. Velocity vectors projected on the slice are shown where gas density is relatively high. The spin direction of galaxy stellar mass projected on the slice is shown with thick arrows, and the sense of the spin perpendicular to the slice is marked with circling arrows. At most places mass flow patterns are very similar between gas and dark matter except in the interior of dense gas clumps \citep[see also][their Fig.~10]{codis12}. On large scales the velocity field is determined by dark matter, the most massive component of the universe, and gas is just following the dark matter velocity field. Gas and stellar mass flows deviate from the dark matter flow only within galaxies.

In the particular region of HR5 shown in Figure~\ref{fig:filaments} a few nonlinear overdensity peaks and filament segments of gas and dark matter appear at $z \gtrsim 8$. When they first form, the proto-galactic density peaks are elongated in alignment with the filaments that they are embedded in. During the period $8 \gtrsim z \gtrsim 6$, however, they turn around and their disks become roughly perpendicular to the filaments. As a result the filament is wrapped around the forming galaxies. In the meanwhile the spin axes of galaxies are roughly aligned with their filament (see the spin directions within and perpendicular to the slice). This phenomenon seems due to the tidal torque exerted to the proto-galaxies by the large-scale gravitational shear field and the corresponding cold gas flow~\citep{Kimm:2011tx,Tillson2015}. 
% The galaxies shown have $M_{\star}= A \sim B \times 10^9 M_{\odot}$ in this period.
This indicates that the galaxies with $M_{\star} \sim 10^9\,\msun$   in the cosmic morning at  $z=6$ may have acquired disk morphology through asymmetric mass infall onto filaments carrying the initial AM. 

After $z=6$ these first galaxies can undergo mass accretion and mergers along filaments and their spins can  swing~\citep[e.g.][]{Bailin2005,aragon-calvo07,Hahn2007}, which weakens the alignment of their spin with the initial AM~\citep{Bett2012,dubois14b,gonzalez17,wang18}. Filamentary segments then merge together~\citep{Cadiou2020} to build thicker filaments or collapse by themselves and form kinematically and morphologically more complex systems of filaments beaded with galaxies ($6 \gtrsim z \gtrsim 4$). 

The fact that the total AM of galaxies has a very tight correlation with their initial AM supports the above scenario, and can also explain why irregulars or spheroids tend to transform back to disks~\citep{welker14}. Even though their stellar component is temporarily perturbed due to encounters and mergers, the matter surrounding them still traces the initial AM and can be accreted consistently along vorticity-rich cold flows, in such a way that the initial disk morphology is recovered. The correlation however gradually weakens with decreasing redshifts. At $z=4.5$, the fraction of galaxies in $\hat{L}\cdot \hat{L}_{\rm tot}>0.8$ is only one third the fraction at $z=6$.

In the meantime, the morphology of the stellar component of galaxies seems quite susceptible to transformation into irregular or spheroidal morphology. This makes the initial correlation between internal kinematics and morphology weaker with cosmic time, and the correlation between initial AM and galaxy orientation becomes almost negligible by $z=6$~\citep[see also][]{Danovich2015,Cadiou2021}. It remains for us to explain the fraction of temporary morphology (i.e. irregular plus spheroid types) of about 1/3 at mass scales $M_{\star} \lesssim 10^{10}\,\msun$ in the cosmological model adopted here. 

\begin{figure}
\centering 
\includegraphics[width=0.45\textwidth]{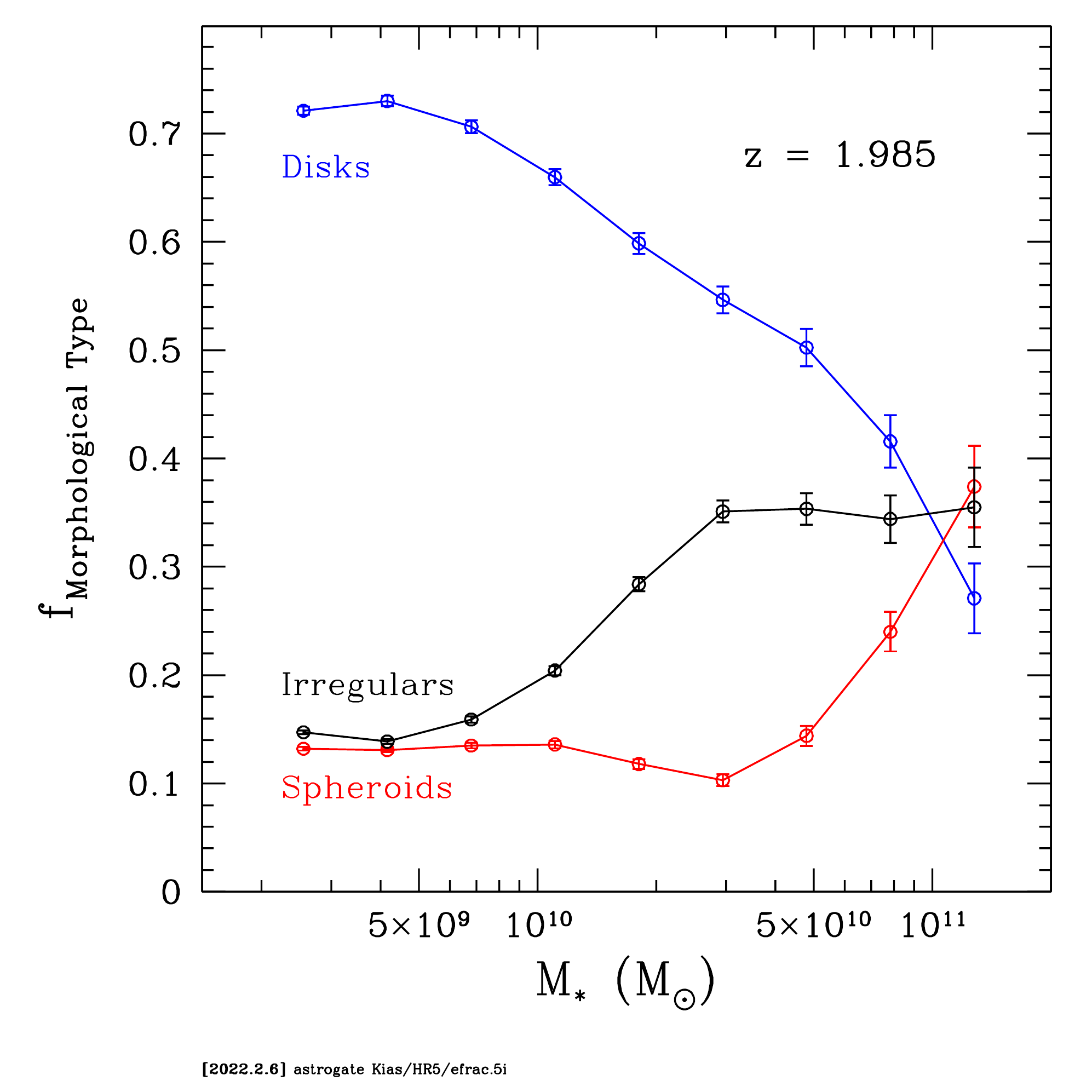}
\caption{Morphological type fractions of galaxies as a function of galaxy stellar mass at $z\approx 2$. The blue, black and red solid lines are for disks, irregulars and spheroids, respectively.}
\label{fig:morphology_mass}
\end{figure}

Let us return to the near independence of morphological type fractions on stellar mass. This finding is true only for the stellar mass range between 2 and $\sim 10\times 10^9\,\msun$ and before the epoch $z \sim 4$. Statistics is not good at higher masses.  %There is an indication that the irregular fraction starts to increase at relatively high masses at $z<5$. 
Even though we focus on galaxy morphology during the cosmic morning before $z=4$,
we have also calculated the fraction of morphological types near the cosmic high noon at $z\approx 2$ to get a hint on what happens to galaxy morphology at later epochs.

Figure~\ref{fig:morphology_mass} shows that the fraction of each morphology is nearly the same as in the cosmic morning for galaxies with $M_{\star} \lesssim 10^{10} \,\msun$, but is radically different at higher mass. In particular, the fraction of spheroids below $M_\star \approx 5 \times 10^{10}\,\msun$ is consistent with the fraction of the GOODS early-type galaxies at $z\approx 1$ reported by \citet{hwang09}. The absolute magnitude limit of $M_{\rm B} = -18.0$ of the GOODS galaxies used in the analysis corresponds to the stellar mass of $M_{\star}\sim3\times10^{10}\,\msun$. This agreement on galaxy morphology fraction at relatively low redshifts between the observation and our simulation provides the basis for our work on morphology of galaxies in the cosmic morning.

A notable feature is a sharp rise of the irregular fraction from $\sim 1$ to $3\times 10^{10} \,\msun$, and about 35\% of massive galaxies above $3 \times 10^{10} \,\msun$ are irregulars. So we expect massive irregular types appear and become frequent as we approach  cosmic noon from  cosmic morning~\citep[see also, e.g.][]{Dubois2016}.

Another dramatic feature in Figure~\ref{fig:morphology_mass} is another steep rise of spheroid fraction at $M_{\star}\sim5 \times 10^{10} \,\msun$. This is probably related with the phenomenon that the irregular fraction starts to rise at smaller mass of $M_{\star}\sim 1 \times 10^{10} \,\msun$. One can interpret the mass scale difference as mass-dependent morphology transformation process from disk morphology via irregular type to spheroid type as galaxies grow in mass. Above the mass scale of $1 \times 10^{11} \,\msun$ disk morphology is now a minor type with its fraction of only about 26\%.  In the forthcoming paper
we will inspect how the fraction of each galaxy morphological type as a function of stellar mass evolves during the cosmic noon (from $z = 4$ to 1.5 as defined by this paper) and afternoon ($z < 1.5$).

\section{Conclusions}

Using our new cosmological simulation, Horizon Run 5 (\hr), we find that the initial physical morphology of the galaxies in the cosmic morning, before $z=4$, is disk-dominant with the S\'{e}rsic index of stellar-mass distribution less than 1.5 for about 2/3 of galaxies. % Besides this we have a number of findings.
The fraction of irregulars or spheroids is about 1/6 each. The fractions are nearly independent of redshift and also of stellar mass up to $\sim 10^{10}\,\msun$. Stellar mass jump events in the cosmic morning do not preferentially drive galaxy morphology toward spheroid shape at the mass scales $M_{\star} \lesssim 10^{10}\,\msun$. However, the galaxies passing through irregular-type phase have a weak tendency to end up having spheroid morphology.

Almost all the first galaxies in the cosmic morning form at the initial matter-density peaks. Higher peaks form galaxies at higher redshifts. For example, the galaxies having $M_{\star} = 2.0 - 2.4 \times 10^9\,\msun$ at a redshift $z$ correspond to the peaks with mean height ${\bar \nu}_{\rm pk} = 0.472 z +1.32 \pm 0.01$ in the initial overdensity field smoothed over $R_G =0.35$ cMpc. That is, galaxies with $M_{\star}\approx 2\times 10^9 M_{\odot}$ at $z=6$ have formed at initially $\sim 4.1\sigma$ density peaks. More massive galaxies at the same epoch are the ones that have formed earlier at higher peaks and grown in mass.
 
Large-scale structures in the universe emerge and grow like cosmic rhizomes as the underlying matter-density fluctuations become non-linear and form associations of galaxies in globally overdense regions and stretch the realm of the world of galaxies into lower-density regions along multiply-connected filaments. The cosmic web structure of large-scale distribution traced by galaxies forms at lower redshifts when most rhizomes percolate.

The initial angular momentum produced by the tidal torque on protogalactic regions is tightly correlated with the internal kinematics and spin parameters of galaxies confirming the validity of tidal torque theory. However, the actual shape of the stellar component of galaxies has almost no correlation with the initial angular momentum. These findings can be reconciled if the initial morphology of all galaxies is disk-like, while the morphology of the stellar component of the first galaxies is set up 
%is quite susceptible of temporal transformation due to various kinds of interaction and evolution.
%The initial angular momentum of the proto-galaxies set up
by the anisotropy of the larger-scale tides which induce coherent cosmic gas inflow. % and cause galaxies to form as disks.

%Since  \hr\, does not  agree with the observed universe in every detail, our results cannot be  considered to be fully quantitatively  accurate.
The validity of our predictions, of course, depends on how accurately \hr\ matches the observed universe.
Since we have studied the physical morphology of galaxies using stellar mass density, our results cannot be directly compared  with the apparent morphology from observations that rely on light distribution.
Nonetheless, we hope that many of our findings be judged in view of upcoming observations of the deep universe by the James Webb Space Telescope.

\section*{acknowledgments}
We thank Katarina Kraljic for helpful comments. CP and JK are supported by KIAS Individual Grants (PG016903, KG039603) at Korea Institute for Advanced Study. JL is supported by the National Research Foundation of Korea (NRF-2021R1C1C2011626). BKG and CGF acknowledge the support of STFC through the University of Hull Consolidated Grant ST/R000840/1, access to {\sc viper}, the University of Hull High Performance Computing Facility, and the European Union’s Horizon 2020 research and innovation programme (ChETEC-INFRA -- Project no. 101008324).
This work benefited from the outstanding support provided by the KISTI National Supercomputing Center and its Nurion Supercomputer through the Grand Challenge Program (KSC-2018-CHA-0003). This research was also partially supported by
the ANR-19-CE31-0017 \href{http://www.secular-evolution.org}{http://www.secular-evolution.org}.
Large data transfer was supported by KREONET, which is managed and operated by KISTI. This work is supported by the Center for Advanced Computation at Korea Institute for Advanced Study. 

\newpage

\appendix
\section{S\'{e}rsic index and morphology of galaxies}\label{sec:sersic_index}

\begin{figure}
\centering 
\includegraphics[width=0.4\textwidth]{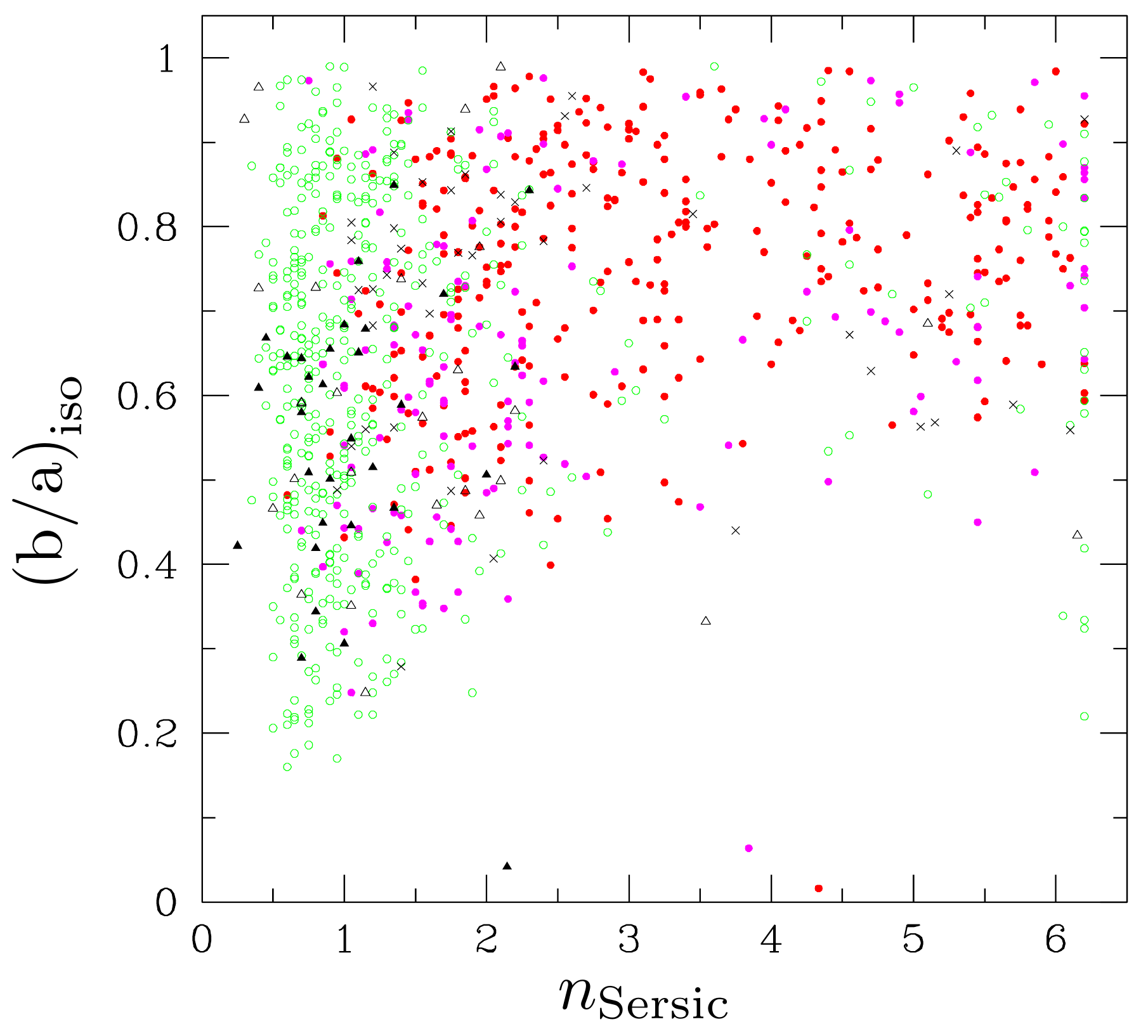}
\caption{Relation between iso-photal axis ratio and S\'{e}rsic index of late-type (green circles), early-type (red or magenta circles), abnormal late-type (triangles), and abnormal early-type (crosses) SDSS galaxies in the training set of \citet{park05}. }
\label{fig:iso_sersic}
\end{figure}

We classify galaxy morphology into disk and spheroid types for symmetric galaxies according to the S\'{e}rsic index $n_{\rm S\acute{e}rsic}$. We determine the S\'{e}rsic index criterion from the morphology training set of SDSS galaxies given by~\cite{park05}. The sample contains visually-identified morphology information of 1982 nearby bright late-type (spirals and irregulars) and early-type (ellipticals and lenticulars) galaxies. We measure the radial surface brightness profile using an ellipse fitting method to account for the inclination effects, and find the best-fit $n_{\rm S\acute{e}rsic}$ between 0.2 and 2 times the Petrosian radius~\citep{park05,park07}. The central region has been discarded to mitigate the effects of finite point spread function of the SDSS survey (about $1.4''$). Figure~\ref{fig:iso_sersic} shows the iso-photal axis ratios and S\'{e}rsic indices of late-type galaxies (green circles and triangles) and early-type SDSS galaxies (red/magenta circles and crosses) in the set. The distribution of $n_{\rm S\acute{e}rsic}$ is continuous, and the transition from late to early type is not abrupt.
However, it can be seen that when galaxies are classified by using $n_{\rm S\acute{e}rsic}$ alone, the threshold of $n_{\rm S\acute{e}rsic} = 1.5$ is a good division criterion.

\bibliographystyle{apj}

\end{document}